\documentclass[%
 reprint,
 amsmath,amssymb,
 aps,
]{revtex4-2}

\usepackage{graphicx}
\graphicspath{ {./figures/} {./figures_numeric} }

\usepackage{dcolumn}
\usepackage{bm}

\usepackage{amssymb}
\usepackage{amsmath}
\usepackage{multirow}
\usepackage{textcomp}
\usepackage{gensymb}
\usepackage{hyperref}

\usepackage{mathMacros}

\usepackage{float}
\usepackage{array}
\newcolumntype{P}[1]{>{\centering\arraybackslash}p{#1}}
\usepackage{caption}
\usepackage{subcaption}
\usepackage[title]{appendix}
\newcommand\parrw{0.25}

\usepackage[normalem]{ ulem }
\usepackage{color}

\begin{document}

\preprint{APS/123-QED}

\title{Experimental and numerical investigation\\ of bubble migration in shear flow:\\ deformability-driven chaining and repulsion}

\author{Blandine Feneuil}
 \affiliation{Department of Mathematics, University of Oslo, Oslo, Norway}
 \affiliation{SINTEF Industry Petroleum, Trondheim, Norway}
 \email{blandine.feneuil@sintef.no}

\author{Kazi Tassawar Iqbal}%
 \affiliation{FLOW and SeRC, KTH Royal Institute of Technology, Stockholm, Sweden  }%

\author{Atle Jensen}%
 \affiliation{Department of Mathematics, University of Oslo, Oslo, Norway  }

\author{Luca Brandt}%
 \affiliation{FLOW and SeRC, KTH Royal Institute of Technology, Stockholm, Sweden}%
 \affiliation{Department of Energy and Process Engineering, NTNU, Trondheim, Norway}%

\author{Outi Tammisola}
 \affiliation{FLOW and SeRC, KTH Royal Institute of Technology, Stockholm, Sweden}

\author{Andreas Carlson}
 \affiliation{Department of Mathematics, University of Oslo,  Oslo, Norway}

\date{\today}

\begin{abstract}
We study the interaction-induced migration of bubbles in shear flow and observe that bubbles suspended in elastoviscoplastic emulsions organise into chains aligned in the flow direction, similarly to particles in viscoelastic fluids. 
To investigate the driving mechanism, we perform experiments and simulations on bubble pairs, using suspending fluids with different rheological properties. First, we notice that, for all fluids, the interaction type depends on the relative position of the bubbles. If they are aligned in the vorticity direction, they repel, if not, they attract each other. The simulations show a similar behavior in Newtonian fluids as in viscoelastic and elastoviscoplastic fluids, as long as the capillary number is sufficiently large. This shows that the interaction-related migration of the bubbles is strongly affected by the bubble deformation. We suggest that the cause of migration is the interaction between the heterogeneous pressure fields around the deformed bubbles, due to capillary pressure.
\end{abstract}

\maketitle

\section{Introduction}

An aerated mortar and a \textit{mousse au chocolat} are just two of many industrial fluids that contain bubbles. These materials are mixed and transported until they are set in place, \ie sprayed on a wall or poured in a pot. During this process, the distribution of the bubbles must remain homogeneous for the materials to fulfil their function, respectively thermal insulation and taste. These two examples illustrate the necessity to control bubble migration during the processing of industrial fluids. These fluids often have complex rheological properties. They are called shear-thinning if the dynamic viscosity decreases when the shear rate increases. In addition, they may display a viscoelastic behavior. Elasticity is characterized by normal stress differences, \ie differences between the diagonal terms of the stress tensor, as well as memory effects. The goal of this paper is to investigate bubble interactions in a sheared fluid, and focus on the effect of the suspending fluid rheological properties on the relative migration of the bubbles.

A neutrally-buoyant solid spherical particle suspended in Newtonian Stokes flow follows the streamlines. However, literature shows that  the reversibility of the trajectory can be broken, even at vanishing Reynolds number, as soon as the system is more complex. Three main causes are relevant to our study: (1) deformability of the suspended object (particle, droplet or bubble), (2) the rheological properties of the fluid, and (3) the interactions between several suspended objects.
 First, a deformable object tends to migrate away from the wall in confined geometries, to regions of zero or minimum shear. Deformable droplets and soft particles have been observed to migrate to the geometry centerline in pressure-driven flow \cite{2016_Villone, 2019_Villone} and in shear flow \cite{1979_Chan, 2014_Villone}. The equilibrium position is affected by the shear-rate heterogeneities: a deformable drop migrates closer to the inner cylinder in a Couette cell \cite{1981_Chan}. Second, in viscoelastic fluids, non-deformable particles move towards the centerline of the geometry or the wall depending on their initial position and size \cite{1979_Chan, 2014_Villone, 2016_Villone, 2019_Villone}.  Particle migration in elastoviscoplastic fluids is similar but more intricate: particles can find equilibrium positions between the centerline and the wall, but they are also able to break the central plug region and migrate to the centerline \cite{Emad_EVPmigration}. Migration of deformable objects in viscoelastic fluids is complex, as it results from both the effect of the rheological properties and the deformability \cite{1979_Chan,1981_Chan,2014_Villone, 2016_Villone, 2019_Villone}. 
 The agglomeration of objects suspended in viscoelastic fluids has been studied extensively in the case of solid particles, for several flow configurations: flow in channels \cite{2018_DelGiudice,2020_DAvino}, sedimenting particles \cite{2002_Daugan,2006_Jie,2018_Chaparian} and simple shear flow \cite{1977_Michele,2001_Lyon,2004_Won,2004_Scirocco, 2010_Pasquino_a, 2010_Pasquino_b,2011_Santos,2014_VanLoon,2016_Jaensson}. In all these studies, particles tended to agglomerate into chains, aligned in the direction of the flow. \textit{Michele \etal} \cite{1977_Michele} first observed long particle chains in viscoelastic fluids in simple shear flow in 1977. They suggested that the chains form due to the normal stress differences, and proposed a criterion for chain formation based on the elasticity parameter defined by:

\begin{equation}
S = \dfrac{N_1}{2 \tau},
\end{equation}
where $\tau$ is the viscous stress and $N_1 = \sigma_{11} - \sigma_{22}$ is the first normal stress difference such that $\sigma_{11}$ is the stress in the flow direction and $\sigma_{22}$ the stress in the shear direction. 
 
Later, several studies have investigated particle alignments in shear flow, and although all agree that elasticity plays a major role in forming chains, they showed that Michele's criterion is not fully valid \cite{2004_Won,2004_Scirocco,2014_VanLoon,2016_Jaensson}. Studies showed that particle alignment could occur in some fluids when $S \sim 1$, while in Boger fluids, \ie elastic fluids with constant viscosity, no alignment was observed up to $S \sim 100$ \cite{2004_Won,2004_Scirocco,2014_VanLoon}. An explanation was proposed by \textit{Van Loon \etal} \cite{2014_VanLoon}: for particle chains to be observed, not only must the particles migrate towards each other, but also, the formed agglomerate must be stable. While the first condition is fulfilled in the presence of normal stress differences \cite{2016_Jaensson}, the stability requires shear-thinning \cite{2014_VanLoon}. In a stable particle chain in shear flow, momentum balance requires that each individual particle in the chain rotates. If not, the chain behaves like a single elongated object, i.e. it tumbles to align in the vorticity direction, eventually leading to the separation of the particles. Shear-thinning is necessary because it allows for the strong shearing of the fluid between neighbor rotating particles \cite{2014_VanLoon}.

Chains of bubbles have not yet been reported in shear flow. Though, it is noteworthy that \textit{Michele \etal} \cite{1977_Michele} mentioned that when they happened to have entrained bubbles in their samples, these bubbles took part in the particle chains. 
Bubble interaction have been investigated in the case of two buoyant bubbles rising in a suspending fluid. Although this configuration is different from shear flow, which is the focus of the present study, some features are worth mentioning. First, attraction between bubbles in a Newtonian fluid has been observed even at vanishing Reynolds numbers (down to 0.005) \cite{1994_Manga}. Secondly, at Reynolds numbers below 10, two types of interactions have been observed. When the bubbles are aligned vertically, they attract, and if they are aligned horizontally, they repel \cite{1996_Katz,2003_Legendre,2011_Velez,2017_Gumulya,2019_Kong}. Finally, these studies evidence the major role of deformability on rising bubble interaction in Newtonian fluids at low Reynolds numbers \cite{1994_Manga, 2017_Gumulya, 2019_Kong}. \textit{Vélez-Cordero \etal} \cite{2011_Velez} also studied the effect of suspending fluid shear-thinning and noted that it increases the attraction in the vertically-aligned case.

The goal of the present study is to investigate the behavior of bubbles in simple shear flow. First, we study experimentally bubble suspensions and observe that bubbles form chains similarly to particles. To understand the mechanisms driving bubble migration, we conduct experiments and simulations on two bubbles. We observe that a pair of bubbles either attract or repel each other depending on their relative positions. By comparing the trajectories for Newtonian, viscoelastic and elastoviscoplastic suspending fluids, we show that the bubble migration is caused mainly by their deformability.

\section{Materials and methods}

\subsection{Material and experimental setup}

\subsubsection{Experimental setup} 

\begin{figure*}
\begin{center}
\includegraphics[height=6cm]{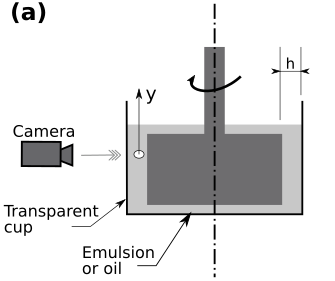}
\includegraphics[height=6cm]{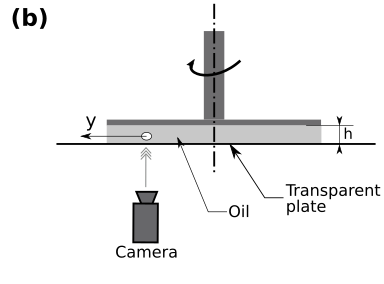} 
\smallbreak
\includegraphics[height=3.5cm]{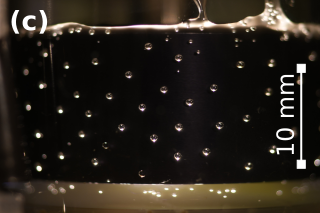} 
\includegraphics[height=3.5cm]{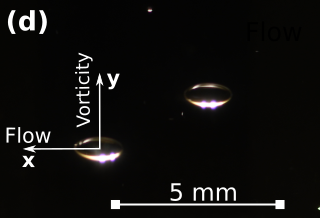} 
\includegraphics[height=3.5cm]{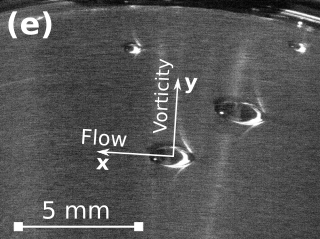} 
\end{center}
\caption{Schematic description of the experimental setup: (a) Couette cell, (b) plate-plate cell. Note that the drawings are not to scale. (c) Example of a bubble suspension in the Couette cell, taken before shearing (emulsion Em3, d = 0.4~mm). 
(d) Example of a bubble pair during shearing in the Couette cell (emulsion Em2, d = 1~mm, $\dot{\gamma} = 5~s^{-1}$). Here, $\Ca \sim 1$ and the bubbles elongate in the direction of the flow. (e) Example of a pair of bubbles in oil V10K in the plate-plate cell. We denote $x$ the flow direction and $y$ the vorticity direction.}
\label{Fig_setup}
\end{figure*}

Most of the experiments are carried out in the Couette cell C-LTD 70/PIV provided by Anton Paar (see Fig. \ref{Fig_setup}(a)). It comprises a 35 mm-diameter cup and a 32 mm-diameter and 16 mm high inner cylinder; the shearing gap is therefore $h = 1.5$~mm. The external wall of the cell is transparent, allowing for the visualization of the bubbles with a camera (Nikon D7500) during the whole duration of the experiment. A light source placed below the Couette cell creates a bright reflection on the bubble surface, which makes the bubbles appear clearly on the camera pictures (see Fig. \ref{Fig_setup}(c) and (d)). Fluid circulation inside the external wall allows for the regulation of temperature in the cell; for all the experiments, the temperature is set to 23\degree C. The picture acquisition rate is between 2 images per second and 1 image/4 s, depending on the imposed shear rate $\dot{\gamma}$. Both the inner and outer wall of the geometry are smooth. We have checked that no major wall slip occurs with the fluids when $\dot{\gamma}\geq 1 s^{-1}$ by comparing the flow curves obtained with the transparent Couette cell to those obtained in a rough plate-plate geometry, and by using Particle Image Velocimetry in the transparent Couette cell \cite{2020_Feneuil}. In the Couette cell, the flow direction is horizontal and the vorticity direction is vertical (Fig. \ref{Fig_setup}(d)).

A few additional experiments have been performed in a plate-plate setup (Fig. \ref{Fig_setup}(b)). The upper plate is a 5-cm diameter smooth steel disk provided by Anton Paar. The bottom plate is a transparent plexiglas plate, allowing for the visualization of the bubbles with a camera placed below the setup. The plate-plate setup allows us to choose the shearing gap h, we have taken h = 1.5 mm. In the plate-plate cell, the vorticity direction is radial (Fig. \ref{Fig_setup}(e)).

Before the experiments, the fluid is placed in the Couette cell or between the plates, and when necessary, sheared at 50~s$^{-1}$ for several minutes to remove previously entrained bubbles. When all the artifact bubbles are removed, we inject bubbles of equal size with a microsyringe. Injected bubble volume $V_b$ is between 0.025~$\mu$L and 0.5 $\mu$L. The equivalent diameter $d$ and radius $R$ are defined as

\begin{equation}
d = 2R = \sqrt[3]{\dfrac{6 V_b}{\pi}}.
\end{equation} 
Therefore, 0.4 mm $\leq d \leq$ 1 mm. The confinement ratio is defined as 

\begin{equation}
\beta =  \dfrac{d}{h}.
\end{equation}
In the Couette cell, the gap size h is constant, we have $0.27 \leq \beta \leq 0.67$. In the plate-plate cell, 1~mm diameter bubbles are used and the gap size h is 1.5~mm, giving a confinement ratio $0.67$. Constant shear rate is imposed and pictures of the bubbles are taken. Afterwards, the pictures are post-processed with Matlab to extract the position of the bubbles. 

Two types of tests are carried out. In the bubble suspension tests, we inject between 100 and 250 bubbles in the Couette cell (see Fig. \ref{Fig_setup}(c)) and we perform a statistical analysis of the bubble positions. First, we observe whether bubbles form chains during shearing. Then, we evaluate the number of chained bubbles as a function of time. 
The shear rate is $ 1~s^{-1} \leq \dot{\gamma} \leq 15~s^{-1}$, and the bubble diameter is between 0.4 and 1 mm. 

Experiments with two bubbles are carried out at $\dot{\gamma} = 5~s^{-1}$ and d = 1~mm. Most of the tests are performed in the Couette cell. In a few additional experiments, we investigate the effect of the curvature of the geometry by using the plate-plate cell (see example of pictures on Fig. \ref{Fig_setup}(d) and (e)). For these experiments, we measure the evolution of the bubble separation distance $\\dy$ in the vorticity direction with time.

\subsubsection{Emulsions}

In order to assess the effect of emulsion rheology on particle chaining, we prepare three different oil-in-water emulsions, labeled as Em1, Em2 and Em3. They contain the same continuous phase (3\% by weight of TTAB surfactant in a 53\% by weight glycerol solution) and dispersed phase (silicone oil V350 from VWR); the continuous and dispersed phase have the same optical index so that the emulsions are transparent. The emulsions differ by the droplet size, which depends on the mixing velocity \cite{2018_Zhang}, and by the amount of continuous phase. 
Oil droplets are obtained by mixing 83.5\% by volume of silicone oil into the continuous phase with a Silverson emulsifier. Em1 is mixed for 100 min at 600 rpm and used without being diluted. To obtain Em2 and Em3, rotation velocity of the mixer is increased from 600 rpm to 2400 rpm by steps of 600 rpm, until total mixing time is 130 min. The obtained mother emulsion is then diluted with continuous phase until the droplet volume fraction is 76.6\% (Em2) and 70.7\% (Em3).

\begin{figure}
\centering
\begin{subfigure}[b]{0.4\textwidth}
\caption{}
\includegraphics[width = 0.8\textwidth]{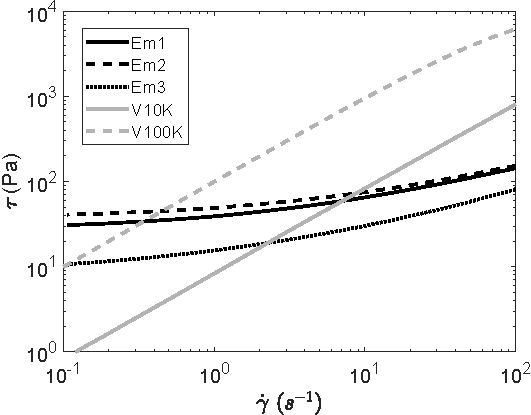}
\label{fig:shear_curve}
\end{subfigure}
\hspace*{0.3cm}
\begin{subfigure}[b]{0.4\textwidth}
\caption{}
\includegraphics[width = 0.85\textwidth]{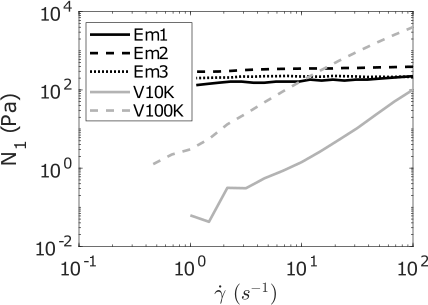}\\
\label{fig:N1_curve}
\end{subfigure}
\begin{subfigure}[b]{0.4\textwidth}
\caption{}
\includegraphics[width = 0.85\textwidth]{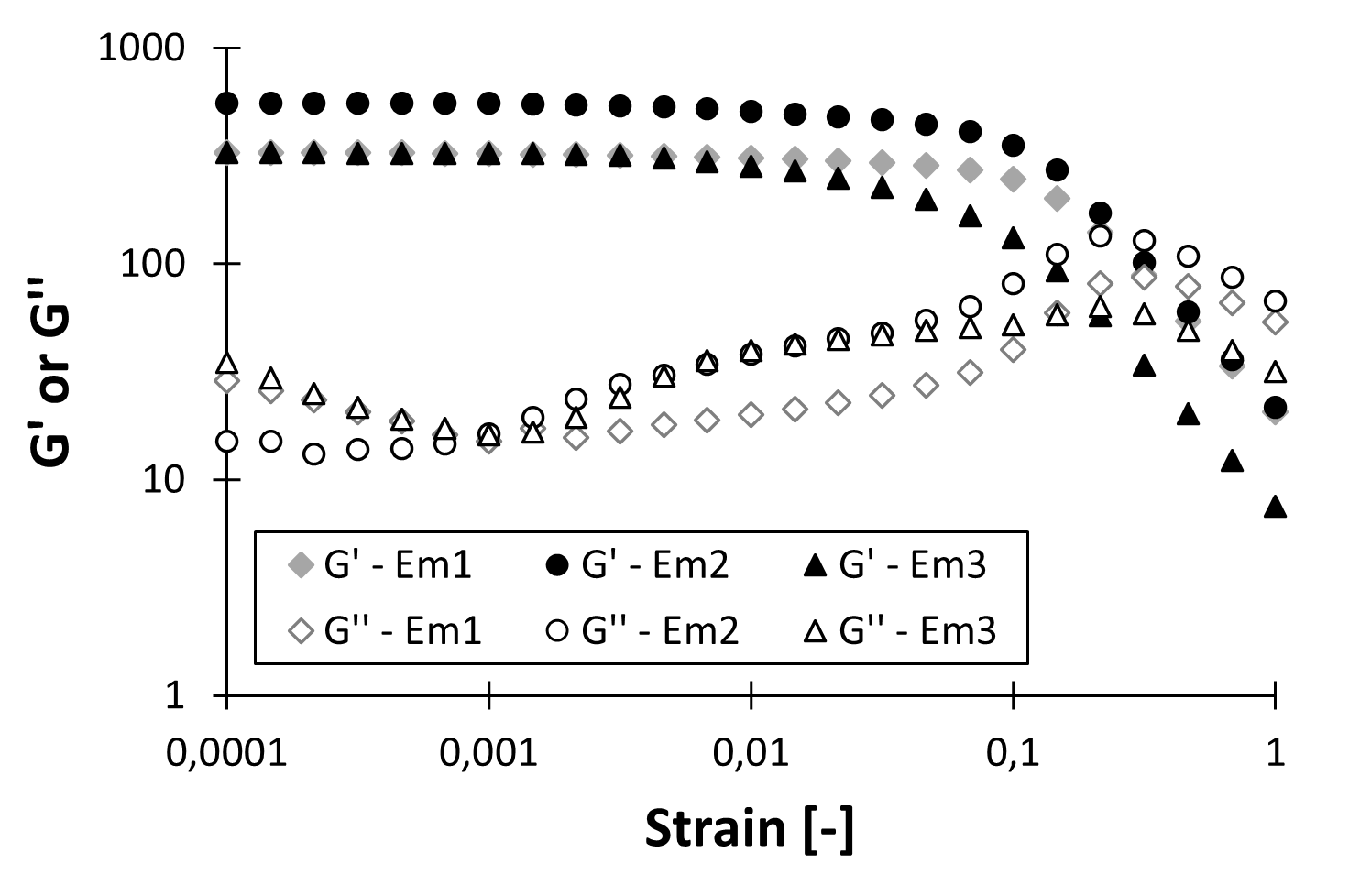}\\
\label{fig:G_curve}
\end{subfigure}
\caption{Rheological properties of the fluids used in the experiments. (a) Viscous shear stress $\tau$ measured by a shear rate ramp. (b) First normal stress difference. In the emulsions (Em1, Em2, Em3), $N_1$ has been measured in a shear rate ramp. In the silicone oils (V10K, V100K), it has been deduced from oscillation tests at angular frequency $\omega$ using the empirical rule $N_1(\dot{\gamma}) \simeq 2 \Gp(\omega)$ \cite{1986_Laun}. (c) Elastic ($G'$) and viscous ($G'$') moduli of the emulsions measured by strain amplitude sweep with oscillation at angular frequency 6.3 rad/s.}
\label{Fig_rheo}
\end{figure}

The emulsions are yield stress fluids, \ie they flow only if the applied stress is above a critical value. This property makes the initialization of the experiments very convenient: the bubbles do not rise as long as the emulsion is at rest, which allows us to inject one by one all the bubbles without them migrating. As soon as the emulsions are set in motion, bubbles rise due to buoyancy. In the Couette cell, we can observe the formation of alignments for a few minutes, until most of the bubbles rise out of the cell. Another advantage of using emulsions is that they contain surfactant, reducing bubble coalescence.

The rheological properties are given for all three emulsions in Fig. \ref{Fig_rheo}. The viscous stress $\tau$ has been measured by applying a shear rate ramp in both a rough plate-plate tool and a smooth cone-tool, which allowed us to check that no wall slip occurs in the cone-plate as soon as $\dot{\gamma} \geq 1~s^{-1}$. The emulsions are strongly shear thinning. The first normal stress difference $N_1$ is deduced from the normal force measured with the cone-plate tool at $\dot{\gamma} \geq 1~s^{-1}$.

By the interpolation of the experimental rheological data, we obtain $\tau(\dot{\gamma})$ and $N_1(\dot{\gamma})$ for all the values of the shear rate. The shear stress is well fitted by a Herschel-Bulkley model: $\tau = \tau_y + k \cdot \dot{\gamma}^n$, where $\tau_y$ is the yield stress of the emulsion, $k$ is the consistency index and $n$ is the flow index. The values of the yield stresses vary from 9 (Em3) to 31 Pa (Em2), while $ n= 0.5$ for all three emulsions. For all emulsions, $N_1$ remains in the same order of magnitude, $N_1 \sim 100~Pa$ with a slight increase in the shear-rate range of interest, i.e. $1~s^{-1} \leq \dot{\gamma}\leq 15~s^{-1}$. We have also performed amplitude sweep oscillation measurements at angular frequency 6.3 rad/s, where we obtain plateau values of the elastic modulus $G'$ between strain amplitudes 0.01\% and 1\%: respectively, 325 Pa for Em1, 550 Pa for Em2 and 320 Pa for Em3 (see Fig. \ref{fig:G_curve}).

With the measured values of $\tau$ and $N_1$, we obtain that the elastic parameter in the emulsions is $1.2 \leq S \leq 7.6$. Due to the high dynamic viscosity $\mu_1(\dot{\gamma})$ of the emulsions, the Reynolds number 
\begin{equation}
\Rey = \dfrac{R^2 \dot{\gamma}\rho_1}{\mu_1(\dot{\gamma})}
\end{equation}
is small, between 10$^{-5}$ and 10$^{-2}$. $\rho_1$ is the suspending fluid density. The rheological properties of the emulsion also affect the deformability of the bubbles: the larger the shear stress, the more likely the bubbles are to deform. Deformability is quantified by the capillary number Ca:

\begin{equation}
\Ca = \dfrac{\tau R}{\Gamma} = \dfrac{\mu_1(\dot{\gamma}) \dot{\gamma} R}{\Gamma},
\end{equation}
where $\Gamma$ is the fluid-air surface tension coefficient. In emulsions, the aqueous continuous phase contains TTAB surfactant above its critical micelle concentration, so $\Gamma = 35$ mN/m \cite{2014_Ducloue} and $ 0.1 \lesssim \Ca \lesssim 1$.

\subsubsection{Silicone oils}

We use two different \textit{Rotational viscosity standard} silicone oils provided by VWR. In the following, V10K will refer to the silicone oil with viscosity 9970 mPa.s at 25 \degree C and V100K will refer to the oil of viscosity 100 075 mPa.s. Their rheological properties are shown in Fig. \ref{Fig_rheo}. V100K is slightly shear thinning above $\dot{\gamma} > 10~s^{-1}$. V10K has a constant viscosity from $\dot{\gamma} = 0.1~s^{-1}$ to $\dot{\gamma} = 100~s^{-1}$. 
The first normal stress differences are very low and could not be deduced from the normal force during the shear-rate ramp experiment. Indeed, the accuracy of the normal force given by the rheometer is a few hundredths of a Newton, corresponding to $N_1 \sim 10~\mathrm{Pa}$. This order of magnitude is exceeded in V10K only when $\dot{\gamma} \gtrsim 100~s^{-1} $ and in V100K when $\dot{\gamma} \gtrsim 10~s^{-1} $. In order to evaluate $N_1$ for the shear rate values of interest, we have performed oscillation tests with increasing oscillation angular frequency $\omega$, at strain amplitude 1\%. 
For polymer melts, an empirical relation between the storage modulus $\Gp$, measured with oscillations of angular frequency $\omega$, and the first normal stress difference states $N_1(\dot{\gamma}) \simeq 2 $G'$(\omega)$ when $\dot{\gamma} = \omega$ \cite{1986_Laun}. $\Gp$ being a shear property, it can be measured with a better accuracy than the normal stress. Using the larger shear rates when $N_1 \gtrsim 10~\mathrm{Pa}$, we have checked that the rule is valid for the silicone oils. $N_1$ down to $\sim$ 1 Pa can therefore be measured with the oscillation tests. The $N_1$ values for the silicone oils are shown in Fig. \ref{Fig_rheo}(b). We observe that $N_1 = \psi_1 \gamma^2$, where the first normal stress coefficient $\psi$ is nearly constant for each silicone oil, $\psi_1 \sim 0.01~Pa.s^2$ for V10K and $\psi \sim 1~Pa.s^2$ for V100K.

Silicone oil V10K was selected because its viscosity is close to the one for the emulsion for our experimental conditions. The surface tension coefficient is $\Gamma = 21$ mN/m. Reynolds and capillary numbers are therefore in the same range as for the emulsions: $\Rey \sim 10^{-3}$ and $\Ca \sim 1$. In V100K, the viscosity is one order of magnitude larger, and $\Rey \sim 10^{-4}$ and $\Ca \sim 10$. Due to very small elastic stresses in the silicone oils, the elastic parameter $S$ is small: 0.1 in V10K and 0.01 in V100K. The rheological properties of the five fluids used in this paper are gathered in Table \ref{table_rheo}, and a summary of the test parameters is given in Table \ref{Table_materials}.

In silicone oil, no surfactant is present, and the bubbles coalesce as soon as they touch. For this reason, silicone oils have been used only in the bubble-pair experiments to observe bubble relative trajectories.

\begin{table}
\begin{tabular}{ c  c c c  c c c }
 & \multicolumn{3}{c}{\textit{Emulsions}} & & \multicolumn{2}{c}{\textit{Silicone oils}}\\
 & Em1 & Em2 & Em3 & & V10K & V100K \\ \hline \hline
$\tau_y $  (Pa) & 27 & 31 & 9 & & 0 & 0\\
$k (Pa.s^{-n})$  & 12 & 11 & 7 & & 8.3 & 95  \\ 
$n$ & 0.5 & 0.5 & 0.5 & & 1 & 1\\ \hline
$N_1$ (Pa) &  $\simeq$ 200 & $\simeq$ 350 & $\simeq$ 200 & & $0.01 \cdot \dot{\gamma}^2$  & $1 \cdot \dot{\gamma}^2$\\
$G'$ (Pa) & 325 & 550 & 320  & & \multicolumn{2}{c}{Not measured}\\ \hline
\end{tabular}
\caption{Summary or the rheological properties of the emulsions and the silicone oils. The viscous stress $\tau$ is fitted for all five fluids by an Herschel-Bulkley model: $\tau = \tau_y + k \cdot \dot{\gamma}^n$, where for the Newtonian silicone oils, $\tau_y$ = 0 and $n = 1$, and the consistency $k$ is equal to the viscosity.}
\label{table_rheo}
\end{table}

\begin{table}

\begin{tabular}{l c c c }
\multicolumn{1}{c}{\large \textbf{(a)}} & \multicolumn{3}{c}{\textbf{Bubble suspensions}}\\
\hline
\hline
 & \multicolumn{3}{c}{\textit{Emulsions}}\\
 & Em1 & Em2 & Em3\\
\hline
\hline
$\Rey $  & \multicolumn{3}{c}{$10^{-5} \rightarrow 10^{-2}$}\\
$\Ca $ & \multicolumn{3}{c}{$0.1 \rightarrow 1$} \\
$\beta$ & \multicolumn{3}{c}{$ 0.27 \rightarrow 0.67$}\\
$S$ & 1.2 $\rightarrow$ 1.7 & 1.9 $\rightarrow$ 3.2 & 2.6 $\rightarrow$ 7.6 \\
\end{tabular}

\medbreak

\begin{tabular}{ c  c c c  c c c}
\multicolumn{1}{c}{\large \textbf{(b)}}  & \multicolumn{6}{c}{\textbf{Bubble pairs}} \\ \hline \hline
 & \multicolumn{3}{c}{\textit{Emulsions}} & & \multicolumn{2}{c}{\textit{Silicone oils}}\\
 & Em1 & Em2 & Em3 & & V10K & V100K \\ \hline \hline
$\Rey $  & \multicolumn{3}{c}{$ \sim 10 ^{-3} $} & & $ \sim 10 ^{-3} $ &  $ \sim 10^{-4}$\\
$\Ca $ & \multicolumn{3}{c}{$\sim$ 1} & & $\sim$ 1 & $\sim$ 10\\ 
$\beta$ & \multicolumn{3}{c}{0.67} & & 0.67 & 0.67\\
$S$ & 1.4 & 2.7 & 4.3 & & $\sim$ 0.01 &  $\sim$ 0.1\\
\end{tabular}

\caption{Summary of the parameters for (a) the bubble suspension tests and (b) the bubble pair experiments. }
\label{Table_materials}
\end{table}

\begin{figure}
\includegraphics[width=.45\textwidth]{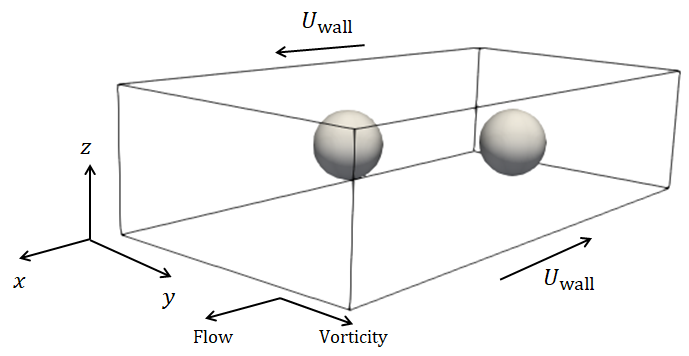}
\caption{Sketch of the computational domain and the coordinate axes. The flow and the vorticity directions are also shown.}
\label{fig:numeric_setup}
\end{figure}

\subsection{Numerical methods}\label{ssec:numerical_method}

\begin{center}
\begin{table*}
\centering
\begin{tabular}{  P{2.5cm}  P{4cm} P{3cm} P{3cm}  }
\hline
\hline
\textbf{Parameter} & \textbf{Definition} & \textbf{Silicone oil} & \textbf{Emulsion} \\ 
\hline
\hline
$\Rey$ & $\rho_1 R^2 \dot{\gamma}/\mu_1$ & 0.05 & 0.05 \\
$\Ca$  & $\mu_1 R\dot{\gamma}/\Gamma$ & 0.5 & $0.2\rightarrow 0.5$ \\ 
$\Wi$  & $\lambda\dot{\gamma}$ & $0.02 \rightarrow 0.1$ & $0.066\rightarrow 9$ \\ 
$S$ & $N_1/2\tau$ & $0.018\rightarrow 0.089$ & $0.12 \rightarrow 6.69$ \\
$\Bing$ & $\tau_y/(\mu_1 \dot{\gamma})$ & 0 & $0 \rightarrow 0.5$\\ 
$\alpha_s$ & $\mu_{s,1}/\mu_1$ & $1/9$ & $1/9$\\
$k_\rho$ & $\rho_1/\rho_2$ & $10$ & $10$\\
$k_\mu$ & $\mu_1/\mu_{2}$ & $10$ & $10$ \\ 
$n$ & - & - & $0.5$\\
$\beta$ & $d/h$ & 0.5 & 0.5\\
\end{tabular}
\caption{Definition of dimensionless numbers and range of values investigated in the numerical simulations. The subscripts $1$ and $2$ denote the suspending and bubble phase respectively.}
\label{tab:params}
\end{table*}
\end{center}

The interaction of bubble pairs migrating under shear flow is also investigated by three-dimensional direct numerical simulations, using the in-house solver and method for elastoviscoplastic fluid simulations presented in \cite{izbassarov2018computational}. Regarding related works of the direct numerical simulation of two fluid-phase systems where one phase is elastoviscoplastic, only a single rising bubble \cite{Tsambubble} and a single droplet in shear flow \cite{EVPdroplet} have been considered so far. Here, the level set method with re-initialization and the continuum surface formulation \cite{brackbill1992continuum} is used to track the interface between the two immiscible fluid phases (bubble and suspending phase). 

Figure \ref{fig:numeric_setup} illustrates the computational setup, along with the coordinate axes and the direction of the wall velocities; the flow and vorticity directions are also shown. A computational domain of size $L_x \times L_y \times L_z = 16R \times 12R \times 4R$ is used for the results in Figs. \ref{fig:oil_attr_intfEvn} and \ref{fig:oil_rep_intfEvn}, and $16R \times 16R \times 4R$ otherwise. A uniform grid spacing of $\Delta x = \Delta y = \Delta z = R/16$ is used. The top wall in Fig. \ref{fig:numeric_setup} moves with a velocity of $U_{\mathrm{wall}}\, \tens{e}_x$ and the bottom wall moves with a velocity of $-U_{\mathrm{wall}}\, \tens{e}_x$ where $U_\mathrm{wall}$ is the magnitude of the wall velocity and $\tens{e}_x$ is the basis vector in the $x$-direction. Periodic boundary conditions are imposed on the streamwise and spanwise boundaries.

The definition of the dimensionless numbers and the reference scales used, the range of dimensionless numbers studied in the numerical simulations - for both emulsion and oil suspending phases - are presented in Table \ref{tab:params}. For the viscosity of the suspending phase, the subscript $s$ denotes the solvent viscosity, $e$ denotes the extra viscosity from the non-Newtonian contribution, and $\mu_1 = \mu_{s,1}+\mu_{e,1}$. Regarding the Weissenberg number $\Wi$ and the elasticity parameter $S$, here we use the same definitions and notation as in \cite{2016_Jaensson}. In general, the experimental and numerical parameters were made as close as possible. In cases where it was not feasible numerically, a compromise between selecting parameters that represent the correct physics and numerical tractability was employed. In particular, the viscosity ratio is significantly lower than that in the experiments. Preliminary tests have however shown that this affects the bubble migration velocity but not the phenomenological observations. Two different methods were used to set the simulation parameters for the emulsion (specifically, the Weissenberg number), for which both set of results are included in \secref{ssec:pair_sims}. In the first method, the Weissenberg number was set in order to match the experimentally measured elasticity parameter, $S$. In the second method, it was set to match the low strain amplitude $\Gp$ from oscillatory shear experiments that were conducted for the emulsions (as in \cite{prudhomme1987response,Tsambubble}, for example).  

Bubbles are modelled as a Newtonian fluid (of viscosity $\mu_2$ and density $\rho_2$). We study bubbles in Newtonian, viscoelastic (VE), and elastoviscoplastic (EVP) suspending phases. The non-Newtonian behaviour of the suspending phase is modeled using a constitutive equation for the VE or EVP stress tensor $\btau^e$. For the oils, the VE Oldroyd-B model \cite{oldroyd1951motion} is used:
\begin{align} 
\label{eq:Oldrd}
\Wi\, \ucd{\btau^e} + \btau^e = 2\tilde{\mu}_{e,1} \tens{D},
\end{align}
where $\btau^e$ is the extra stress tensor due to the non-Newtonian material, and $\tens{D} \equiv \left(\nabla\tens{u} + \trans{\nabla\tens{u}}\right)/2$. $\tilde{(\bullet)}$ denotes material properties normalized with those of the suspending phase. $\ucd{(\bullet)}$ here denotes the upper-convected derivative, 

\begin{align}
\label{eq:ucd}
\ucd{\btau^e} \equiv \pddt{\btau^e} + \advec{\btau^e} - (\grad{\tens{u}})\,\btau^e - \btau^e (\trans{\grad{\tens{u}}}).
\end{align}

To model the emulsion as a shear-thinning, EVP material, the Saramito-Herschel Bulkley constitutive equation \cite{saramito2009new} is used
\begin{align} 
\label{eq:SHB}
\Wi\, \ucd{\btau^e} + F\btau^e = 2\tilde{\mu}_{e,1} \tens{D},
\end{align}
where $F = \max \left(0,\frac{|\btau^e_d| - \Bing}{(2\tilde{\mu}_{e,1})^{1-n}|\btau^e_d|^n }\right)^{1/n}$, $\Wi$ is the Weissenberg number, $\Bing$ is the Bingham number, $n$ is the flow index such that $0 < n < 1$ corresponds to shear-thinning behaviour, and $|\btau^e_d|\equiv \sqrt{\btau^e_d : \btau^e_d /2}$ where $\btau^e_d = \btau^e - (\mathrm{tr}\, \btau^e/\trace{\tens{I}})\, \tens{I}$ is the extra stress deviator tensor (here, $\tens{I}$ is the identity tensor). $F$ is introduced by \textit{Saramito} in Eq. \eqref{eq:SHB} to model the yielding behaviour of the material such that when $F = 0$, the material deforms as a Kelvin-Voigt solid, and when $F > 0$, the material flows as an elastoviscoplastic extension of the Herschel-Bulkley model \cite{saramito2009new}.
To simulate the high Weissenberg numbers at which the emulsions were experimentally studied, we used the log conformation formulation \cite{fattal2004constitutive} to alleviate the well-known high Weissenberg number problem. In this formulation, an evolution equation for the logarithm of the conformation tensor $\bPsi = \log{\tens{C}}$ (where $\tens{C}$ is the conformation tensor) is solved. The velocity gradient $\nabla \tens{u}$ is decomposed into two anti-symmetric tensors $\bgreek{\Omega}$ and $\tens{N}$, and a symmetric tensor $\tens{B}$ which commutes the conformation tensor. The Saramito Herschel-Bulkley constitutive equation in the log conformation tensor formulation
reads 
\begin{align} 
\label{eq:SHBlog} 
\pddt{\bPsi} + \advec{\bPsi} - (\bgreek{\Omega}\bPsi - \bPsi\bgreek{\Omega}) - 2\tens{B} = \frac{F}{\Wi} \left(\exp{(-\bPsi)} - \tens{I}\right).
\end{align}

The flow inside both the bubbles and in the suspending phase is governed by the incompressible Navier-Stokes equations:
\begin{align} 
\label{eq:continuity}
\nabla\cdot\tens{u} = 0,
\end{align}
\begin{align} 
\label{eq:NS}
\tilde{\rho}\Rey \left(\pddt{\tens{u}} + (\advec{} ) \tens{u}\right) = -\grad{p} + \nabla\cdot\btau + \frac{1}{\Ca} \kappa \delta (\phi) \,\tens{n}.
\end{align}
Here, $\tens{u}$ denotes the velocity vector, $p$ the pressure, $\tilde{\rho}$ the normalized density, and $\Rey$ the Reynolds number (see Table \ref{tab:params}). Furthermore, $\Ca^{-1} \kappa \delta (\phi)\tens{n}$ is the surface tension represented as a body force \cite{brackbill1992continuum}, where $\kappa$ is the interface curvature, $\tens{n}$ is the unit normal vector to the interface, and $\delta(\phi)$ is a regularized delta function of the level set function $\phi$. $\btau = 2\tilde{\mu}_{N}\tens{D} + \btau^e$ denotes the sum of the purely viscous component of either phase (\ie the viscous stress tensor of the bubble phase and the solvent contribution in the suspending phase), and the extra stress tensor $\btau^e$ from the VE/EVP phase. Note that the effect of buoyancy is neglected in the numerical simulations; we have simulated some cases with the same buoyancy as in the experiment (\ie same Galilei number) and found there is negligible difference in the relative trajectories of the bubbles when buoyancy's effect is included.

Finally, the level set function $\phi$ is advected by the flow by solving:
\begin{align} 
\label{eq:LS}
\pddt{\phi} + \advec{\phi} = 0.
\end{align}
The signed distance function $\phi$ is defined such that $\phi > 0$ in the suspending phase, $\phi < 0$ in the bubble phase, and $\phi = 0$ at the interface. Re-initialization of $\phi$ is employed, but for brevity we do not include the details of the numerics here. 
Finally, we note that in this numerical study, each bubble was required to be individually identified in order to track their centers of mass to analyze their trajectories, thus two level set functions $\phi_1$ and $\phi_2$ were used to track each interface for the bubble-pair.

Eqs. \eqref{eq:Oldrd}, \eqref{eq:SHBlog}, \eqref{eq:continuity}, \eqref{eq:NS}, are numerically discretized by a second-order finite difference scheme to compute the spatial derivatives. The only exception is the advective term of Eqs. \eqref{eq:Oldrd} and \eqref{eq:SHBlog}, for which a fifth-order weighted essentially non-oscillatory (WENO) scheme is used. For the advective term of Eq. \eqref{eq:LS}, a fifth-order high-order upstream-central scheme is used. A second-order, two-step Adams-Bashforth method is used for the time integration of Eqs. \eqref{eq:Oldrd}, \eqref{eq:SHBlog}, and \eqref{eq:NS}, and a total-variation-diminishing, third-order Runge-Kutta scheme for Eq. \eqref{eq:LS}. The method in \textit{Dodd \& Ferrante} \cite{dodd2014fast} is used to obtain a constant-coefficient Poisson equation for the pressure, which is then solved directly with an FFT-based solver. An extensively validated in-house code is used to perform the simulations, and the interested reader is referred to  \cite{ge2018efficient, izbassarov2018computational} for further details of the numerics, and a large number of validation cases.   

\section{Results}

\begin{figure*}
\begin{center}
\begin{tabular}{l  l  l}
\textbf{(a) Chains}  &  \textbf{(b) Distant alignments} &  \textbf{(c)}\\
\includegraphics[height=4.6cm]{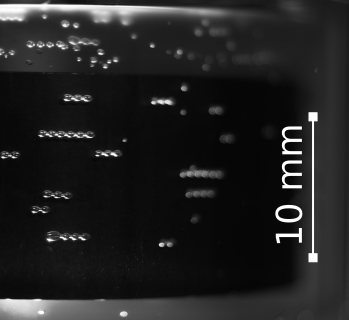} & 
\includegraphics[height=4.6cm]{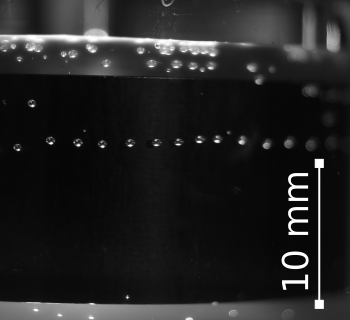} & 
\includegraphics[height=4.6cm]{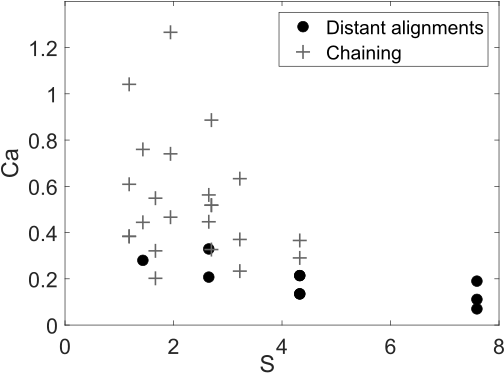}
\\
\end{tabular}
\end{center}
\caption{Pictures of the final configurations observed after shearing bubble suspensions in emulsions: (a) chains and (b) distant alignments. (c) Observed final configurations for all tests as a function of the elasticity parameter S and the capillary number Ca.}
\label{Fig_configurations}
\end{figure*}

\subsection{Experiments on bubble suspension}
\label{section_suspensions}

\subsubsection{Final configurations}

We start by looking at experiments on bubbles suspended in emulsions. After shearing the suspension for several minutes, we observe in all the tests that the bubbles align in the direction of the flow. We distinguish two final configurations, as illustrated in Fig. \ref{Fig_configurations}. In most tests, chain-like structures are formed, where the bubbles come very close to each other. On the pictures, the bubbles appear in contact. 1-mm diameter bubbles in Em1, after coming in contact, tended to coalesce in spite of the high amount of surfactant in the emulsion. In Em2 and Em3, much less coalescence was observed. For the smaller bubbles, the chains sometimes agglomerated into 2D crystal like structures. The 2D structures always formed after the chains by the agglomeration of several chains. Therefore, we have chosen to include the 2D agglomerates within the "chain" configuration.

In a few tests, the final distance between the bubbles in the alignments was several times the bubble diameter. These distant alignments were observed mostly with the smaller bubbles (d $\lesssim$ 0.6 mm) in emulsion Em3, and with the smaller shear rates. Em3 has a lower viscosity $\mu$ than the other two, i.e. the viscous stress $\tau$ is smaller. Distant alignment has been observed before for particles \cite{2004_Won, 2004_Scirocco}. As the duration of our experiments is limited to a few minutes due to bubble rise in the Couette cell, it could be argued that the distant alignments are just a temporary state before chaining. However, the constant inter-bubble distance in these alignments seems to show that the alignments have reached an equilibrium. In the case of particles, \textit{Won \etal} \cite{2004_Won} have checked in their experiments that close-particle chains can be separated into distant alignments by decreasing the shear rate, and concluded that the distant alignments are a dynamic equilibrium state.

In Fig. \ref{Fig_configurations}(c), we show the capillary number and the elastic parameter for all the experiments. Chaining is obtained for $ S = 1 \rightarrow 4$ and distant alignments, for $S = 1 \rightarrow 8$. This shows that, similarly to particles, no chaining criterion can be defined as a critical value of S \cite{2004_Won,2004_Scirocco,2014_VanLoon,2016_Jaensson}. We observe that the distant alignments occur for the smaller capillary numbers, $\Ca < 0.4$, which suggests that bubble deformability enhances chaining. However, in the range $0.2 < Ca < 0.4$, we observe distant alignments for some tests and chaining for others. This shows that bubble deformability does not alone govern the final configuration. Other parameters main play a role such as the fluid shear-thinning behavior and the confinement \cite{2014_VanLoon, 2010_Pasquino_b}.

\subsubsection{Kinetics of chain formation}

\begin{figure}
\textbf{(a)} 

\includegraphics[width = 7cm]{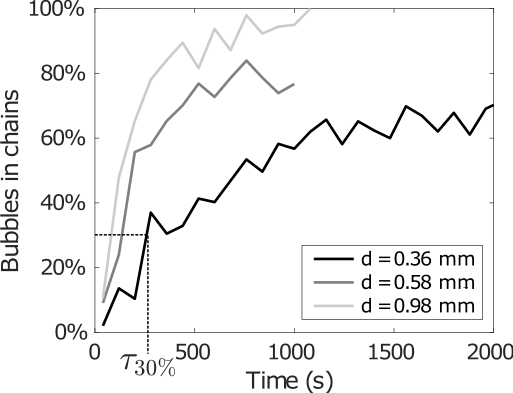} 

\textbf{(b)}

\includegraphics[width = 7cm]{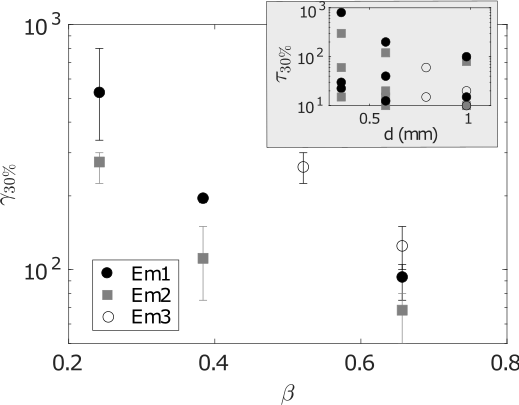}
\caption{(a) Proportion of bubbles in chains as a function of time (Em2, $\dot{\gamma} = 1~s^{-1}$) (b)Strain $\gamma_{30\%}$, reached when 30\% of the bubbles are in chains, as a function of the confinement $\beta$. The error bars indicate the lower and the larger value obtained for the three different shear rates. Inset: Chaining time $\tau_{30\%}$ as a function of the bubble diameter $d$.}
\label{Fig_ChainingTimes}
\end{figure}

Next, we investigate the chain formation process in closer detail. By image analysis, we count the number of bubbles which have a close neighbor. Examples of the obtained curves are given in Fig. \ref{Fig_ChainingTimes}(a). The proportion of bubbles in chains increases just after the start of shearing, then reaches a plateau. The value of the plateau could not be evaluated in Em1 because of the coalescence of the bubbles in the chains. The results obtained with Em2 seem to show that the plateau value depends on the bubble size at $\dot{\gamma} = 1~s^{-1}$. When d = 0.36 mm, about 30\% of the bubbles remain isolated, whereas in the case of the larger bubbles (d = 0.98 mm), very few isolated bubbles are left.  Besides, our results do not show any notable effect of the shear rate on final chaining level of the bubbles (curves not shown here).

From the chaining curves as in Fig. \ref{Fig_ChainingTimes}(a), we identify a characteristic chaining time: $\tau_{30\%}$ is defined as the time for 30\% of the bubbles to get at least one close neighbor. $\tau_{30\%}$ can be also measured in Em1 because coalescence of chained bubbles begins later. The initial number of bubbles in each experiment is between 100 and 250. For experimental reasons, the total number of bubbles tended to increase with the size of the bubbles. In Fig. 5(a), the fastest chaining and the higher chaining plateau are obtained for the largest bubbles even if it is the case where less bubbles are present, which supports the observation that chaining is promoted when the bubble size increases.  One test (Em1, $\dot{\gamma}= 15~s^{-1}$, d = 0.36 mm) has been performed with two different bubble contents (156 and 250), in order to check if the results where affected by the volume concentration of bubbles. We did not measure any difference in the chaining time and the plateau value within the accuracy of the experiments.
Raw chaining times are given in the inset of Fig. \ref{Fig_ChainingTimes}(b) as a function of the bubble size. For each given bubble size and emulsion, three values of the shear rate are investigated, from 1~s$^{-1}$ to 15~s$^{-1}$ and the obtained values for the chaining times range over two orders of magnitude, from 10 to 1000~s. After shearing for time  $\tau_{30\%}$, the accumulated strain in the fluid is $\gamma_{30\%} = \dot{\gamma} \tau_{30\%}$. This strain is plotted for all three emulsions in Fig. \ref{Fig_ChainingTimes}(b):  for a given fluid and bubble size, $\gamma_{30\%}$ does not depend on the shear rate. The chaining dynamics depends therefore on the strain, independently on the velocity. This is similar to observations on particles \cite{2004_Scirocco}. This result is not surprising, as the strain indicates the average travel distance of the particles or bubbles in the fluid: the number of bubble-bubble opportunities to interact is dictated by the strain.

In addition, in each emulsion, $\gamma_{30\%}$ decreases with the bubble size. This can result from two effects: the deformability and the confinement. The larger bubbles are more deformable, and as discussed in the previous paragraph, and the deformability is seen to promote bubbles chaining. Secondly, increasing the confinement ratio increases the probability for the bubbles to meet. Literature on particle chains shows that chaining is promoted by confinement \cite{2004_Scirocco, 2010_Pasquino_b,2013_Pasquino}. Small non-deformable particles in shear flow tend to migrate to the walls in a non-Newtonian fluid \cite{2014_Villone, 2014_Pasquino}, leading to an increase of the particle concentration near the walls, and to an increased probability for particles to meet and interact \cite{2013_Pasquino,2014_VanLoon}. In contrast to particles, the bubbles migrate toward the center of the gap, as we have observed with a camera located below the Couette cell. However, we expect, similarly to particles, that the chaining velocity is enhanced by the increasing probability for bubbles to interact.

The last observation from Fig. \ref{Fig_ChainingTimes}(b) concerns the rheological properties of the emulsion. Surprisingly, the longest chaining times are obtained in Em3, where the viscous stress is low and $ S\geq 2.6$. The elasticity parameter $S$ can therefore not account for the increased chaining time.

To summarize, the behavior of bubble suspensions in emulsions has many similarities with studies on particles in viscoelastic fluids. Two configurations have been observed: chaining with bubbles in contact, and alignment with a larger inter-bubble distance. The chaining kinetics scale with the strain and are enhanced by the confinement. However, the deformability of the bubbles also plays a major role in the interaction, by governing the final configurations and affecting the dynamics. In order to investigate the mechanisms leading to bubble interaction, we focus in the next section on bubble pairs, first experimentally, then numerically.

\subsection{Bubble-pair experiments}\label{ssec:pair_exps}

All the experiments presented in this section are performed at constant shear rate ($\dot{\gamma} = 5~s^{-1}$) and constant bubble size ($d = 1~\mathrm{mm}$).

\subsubsection{Reference tests on isolated bubbles}

First, the trajectory of an isolated bubble is observed in each fluid in the Couette cell, where isolated bubbles rise due to buoyancy. Each test is repeated four times to check that the experimental conditions were reproducible. Examples of trajectories of isolated bubbles are indicated by black lines in Figs. \ref{Fig_trajectories_attraction} and \ref{Fig_trajectories_repulsion}. The bubble rising velocity does not depend on the initial position of the bubble in the gap, i.e. whether it is closer to the rotating cylinder or to the external wall. This ensures that when two bubbles are placed in the cell, a change of their vertical separation distance $\dy$ is a consequence of the interaction between the bubbles. Note that, on the other hand, change of the longitudinal distance $\dx$ may result from the different bubble position in the gap. Therefore, we choose to analyze only the distance in the vorticity direction $\dy$, to understand the interaction between the bubbles.

\subsubsection{Types of bubble-pair interaction}

\begin{figure}
\begin{subfigure}[b]{0.46\textwidth}
\caption{}
\includegraphics[width = 0.9\textwidth]{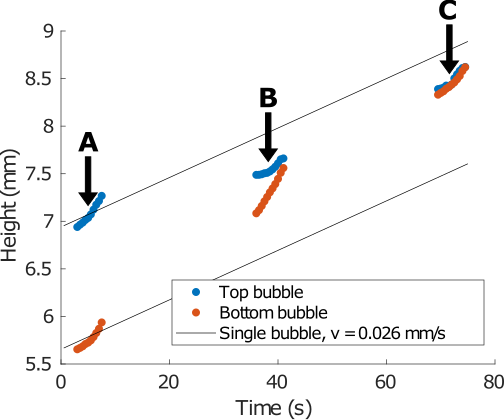} 
\end{subfigure}
\hspace*{0.3cm}
\begin{subfigure}[b]{0.46\textwidth}
\caption{}
\begin{tabular}{l l}
\textbf{A} & \textbf{B }\\
\includegraphics[width = 4 cm]{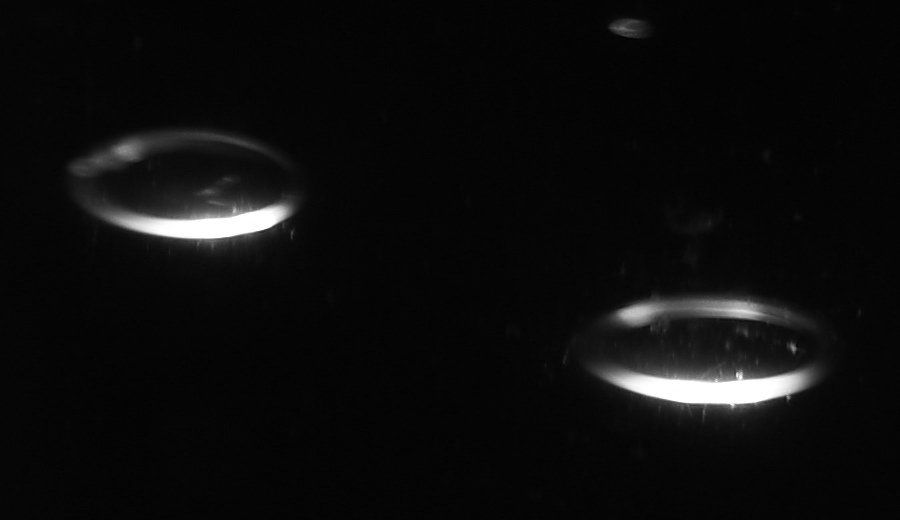} &
\includegraphics[width = 4 cm]{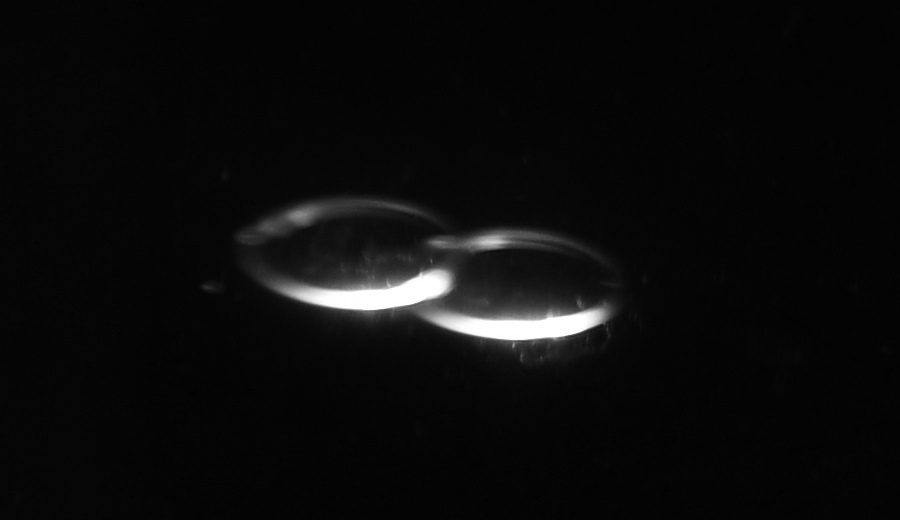}\\
\textbf{C} & \\
\includegraphics[width = 4 cm]{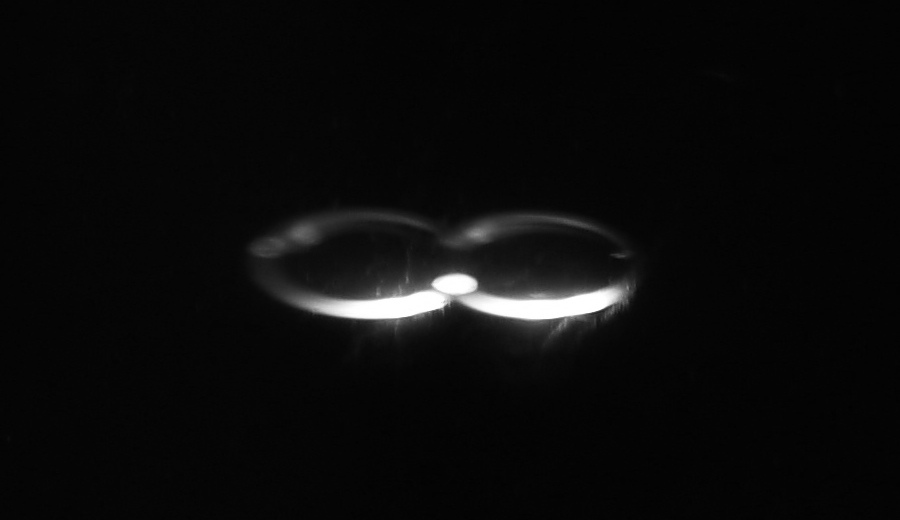}&
\includegraphics[width = 4 cm]{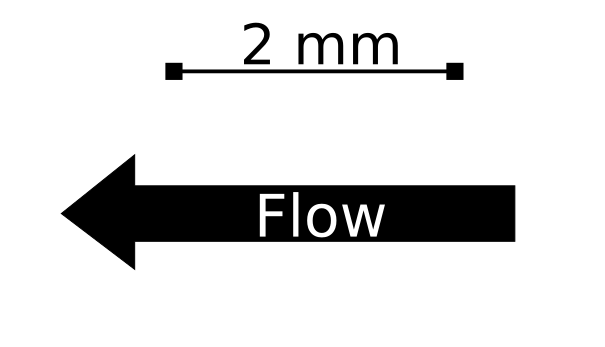}\\
\end{tabular}
\end{subfigure}
\caption{(a) Two bubbles in EM2 interacting in shear flow, where the position of the center of mass of each bubble is tracked in time. The black lines indicate the trajectories of the bubbles if they did not interact. (b) Pictures of the bubbles at three different times indicated by the arrows are shown to illustrate the chaining process.}
\label{Fig_trajectories_attraction}
\end{figure}

When two bubbles are placed in the Couette cell, the rising velocity of each of them differs from the rising velocity of one single bubble. Fig. \ref{Fig_trajectories_attraction}(a) shows an example of the relative displacement of two bubbles with time. The upper bubble rises slower, and the lower bubble rises faster than an isolated bubble. For this test, the bubbles are initially not aligned vertically (see picture A), but one bubble is behind the other in the direction of the flow. After one rotation in the cell, the bubbles have come closer (picture B) and after the second rotation (picture C), they form a stable chain aligned with the flow.

Attraction between bubbles has been observed in all three emulsions and for both oils. Even if $\dy$ is initially three times the bubble diameter, it decreases until the bubbles are chained in the direction of the flow. Distant alignments are never observed in these cases, which, for the bubble size studied here, is in agreement with the results in paragraph \ref{section_suspensions}. In oils, the bubbles coalesce as soon as they touch. In Em1, bubbles tended to coalesce after several seconds in contact. In Em2 and Em3, the doublets were stable for several minutes.

\begin{figure}
\begin{subfigure}[b]{0.46\textwidth}
\caption{}
\includegraphics[width = 0.9\textwidth]{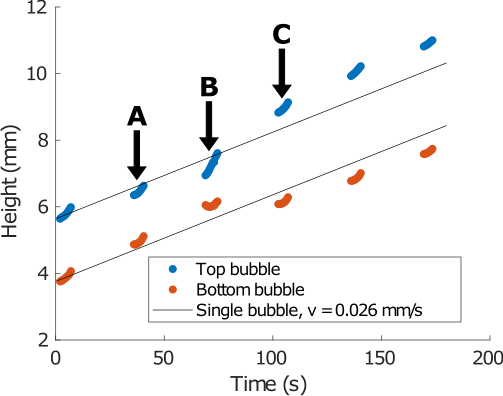} 
\end{subfigure}
\hspace*{0.3cm}
\begin{subfigure}[b]{0.46\textwidth}
\caption{}
\begin{tabular}{l l}
\textbf{A} & \textbf{B}\\
\includegraphics[width = 0.45\textwidth]{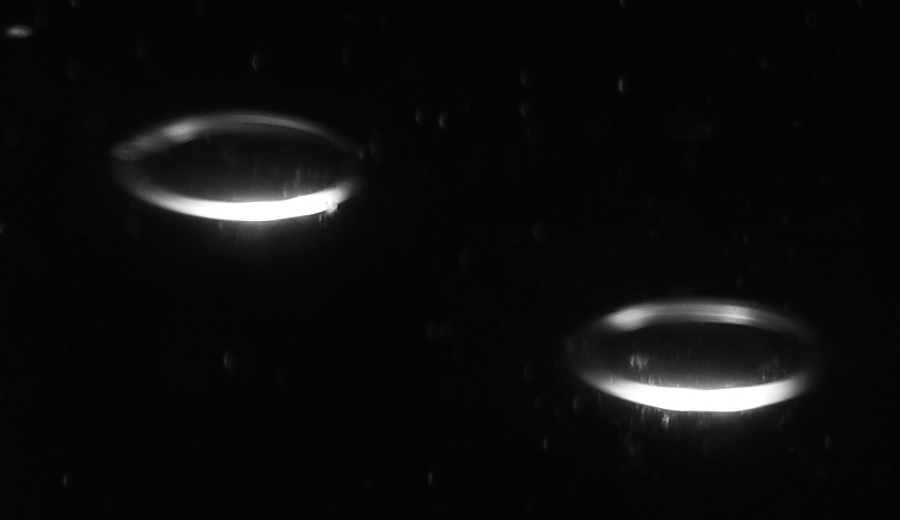} &
\includegraphics[width = 0.45\textwidth]{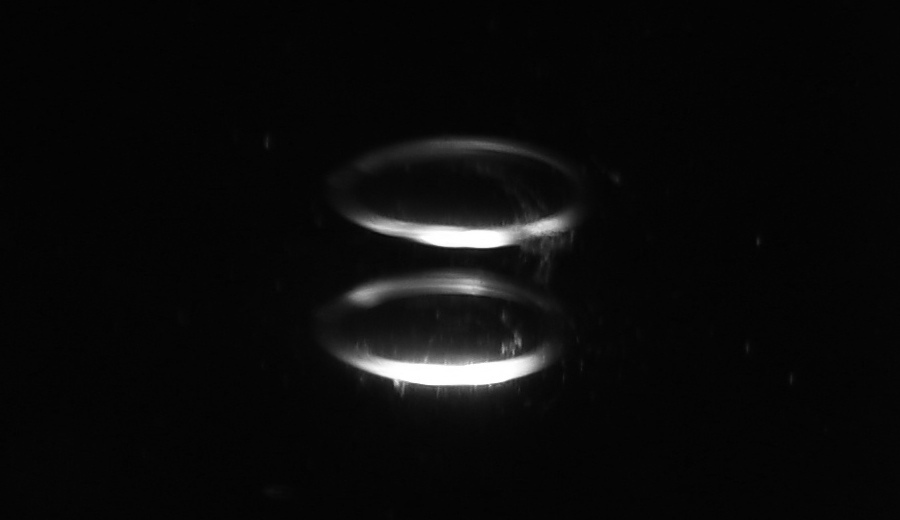}\\
\textbf{C} & \\
\includegraphics[width = 0.45\textwidth]{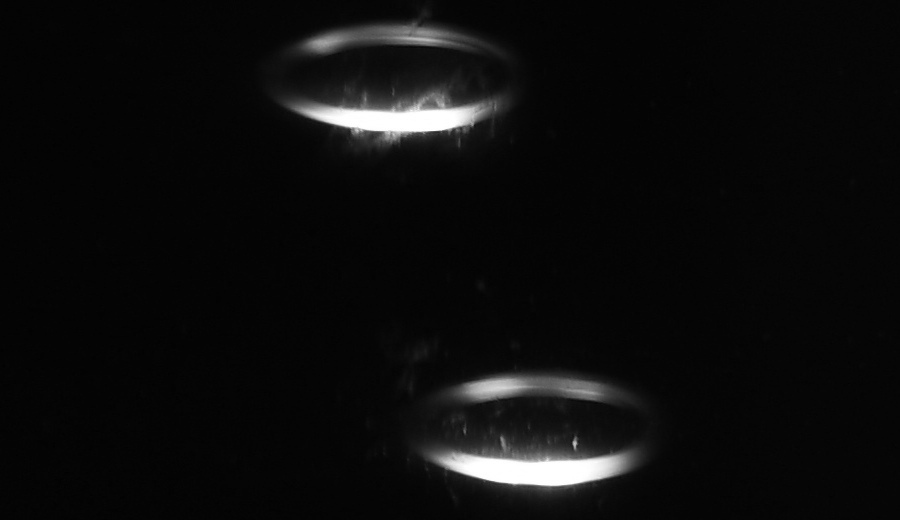}&
\includegraphics[width = 0.45\textwidth]{9_scale.png}\\
\end{tabular}
\end{subfigure}
\caption{(a) Repulsion between two bubbles in Em2 in shear flow, where we track their center of mass with time. The black lines indicate the trajectories of the bubbles if they did not interact. Initially, the bubbles are not aligned in the vorticity direction and attract each other (arrow A). As soon as they become aligned in the vorticity direction (arrow B), they repel each other. (b) A-C show three images at different times as the bubble repel each other.}
\label{Fig_trajectories_repulsion}
\end{figure}

In the test presented in Fig. \ref{Fig_trajectories_repulsion}, bubbles also initially attract and their vertical distance decreases. Though, in the position B, when the bubbles become aligned vertically, the interaction between them becomes repulsive. Repulsion has been observed in multiple experiments in each of the different fluids. When it happens, the bubbles never chain. They can either remain aligned in the vorticity direction, with $\dy/d \leq$ 4, or fully separate. It is worth mentioning that even if repulsion prevents the chaining of bubbles in these experiments, where only two bubbles are involved, it probably enhances chaining in bubble suspensions. Indeed, repulsion of two vertically aligned bubbles in the Couette cell modifies their trajectory and increases the probability for each of them to interact with other bubbles.

\subsubsection{Migration velocity}

In Fig. \ref{Fig_trajectories_repulsion}, the bubbles start repelling each other when they are aligned vertically in the Couette cell, i.e. they are aligned in the vorticity direction. We have checked the consistency of this observation in all the fluids by measuring the direction and the magnitude of the cross-streamline migration velocity of the bubbles. The relative position can be determined at each rotation of the cell, when bubbles are in the field of view of the lateral camera. We determine for each rotation the average relative position:
\begin{equation}
\dx_{av} = \dfrac{\dx_1 + \dx_2}{2} \text{\hspace*{0.5cm} and \hspace*{0.5cm}} 
\dy_{av} = \dfrac{\dy_1 + \dy_2}{2}.
\end{equation}
where $(\dx_1,\dy_1)$ is the relative position of the bubbles before rotation, and $(\dx_2,\dy_2)$ is the relative position of the bubbles after one rotation. The relative velocity $v_y$ in the direction of the vorticity is 
\begin{equation}
v_y = \dfrac{\dy_2-\dy_1}{\Delta t},
\end{equation}
with $\Delta t$ the duration of one rotation. $v_y > 0$ indicates that the bubbles repel each other, and $v_y < 0$, that the bubbles attract each other. To facilitate further analysis and discussion on the data, we interpolate the experimental points to create a velocity map. Interpolation is carried out as follows: for each point A of coordinates ($\dx_{av}/d$, $\dy_{av}/d$), we calculate the distances to the experimental points $A_i$ of coordinates ($\dx_{av,i}/d$, $\dy_{av,i}/d$) and velocity $v_{y,i}$:

\begin{equation}
|A - A_i| = \sqrt{\left(\dfrac{\dx_{av}}{d} - \dfrac{\dx_{av,i}}{d}\right)^2 + \left(\dfrac{\dy_{av}}{d} - \dfrac{\dy_{av,i}}{d}\right)^2 }.
\end{equation}

The interpolated velocity at point A is the weighted average of the experimental velocities in the neighborhood of A:

\begin{equation}
v_y(A) = \dfrac{ \sum\limits_{|A - A_i|\leq 1} v_{y,i} |A - A_i|^{-1}}{\sum\limits_{|A - A_i|\leq 1} |A - A_i|^{-1}}.
\end{equation}

\begin{figure}
\begin{subfigure}[b]{0.46\textwidth}
\caption{}
\includegraphics[height=5cm]{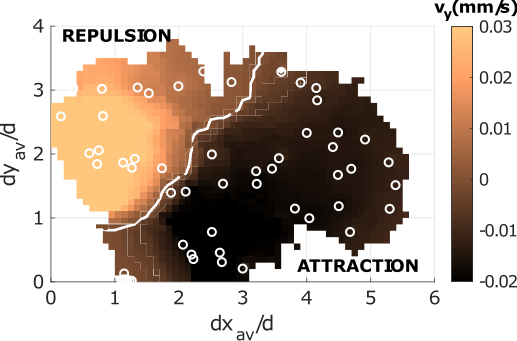}
\end{subfigure}
\hspace*{0.3cm}
\begin{subfigure}[b]{0.46\textwidth}
\caption{}
\includegraphics[height=5cm]{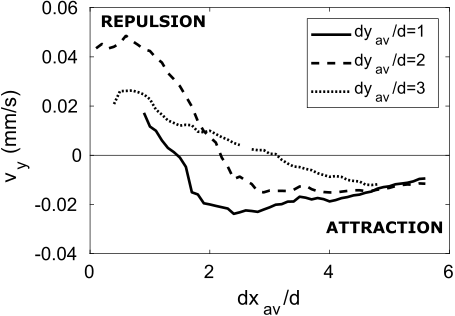}
\end{subfigure}
\caption{Relative velocity of the bubbles in the vorticity direction $v_y$, measured in Em2 in the Couette cell. A positive velocity means that bubbles are repelling each other, and a negative velocity that they are attracting  each other. (a) Velocity map after interpolation as a function of the relative position of the bubbles. The scale bar on the side gives $v_y$ in mm/s. The white line indicates the contour $v_y = 0$ and the white circles indicate the experimental points which have been interpolated to obtain the velocity map (b) Velocity for three different distances $\dy_{av}$.}
\label{Fig_Velocities_Em2}
\end{figure}

\begin{figure}[t]
\begin{subfigure}[b]{0.46\textwidth}
\caption{}
\includegraphics[height=5cm]{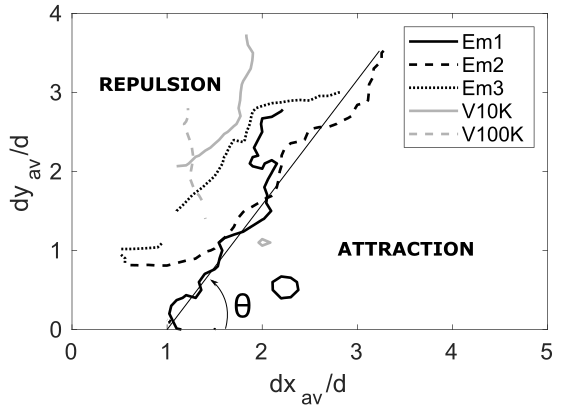}
\end{subfigure}
\hspace*{0.3cm}
\begin{subfigure}[b]{0.46\textwidth}
\caption{}
\includegraphics[height=4.5cm]{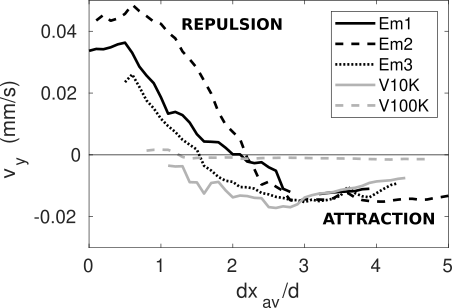}
\end{subfigure}
\caption{(a) Comparison of the location of the contour $v_y = 0$, between the repulsion and attraction regions, for all fluids. The contours make roughly and angle $\theta \simeq 60 \degree$ with the flow direction. (b) $v_y$ plots for all the fluids at $\dy_{av} = 2 d$.}
\label{Fig_Contour_all}
\end{figure}

The example for Em2  is given in Fig. \ref{Fig_Velocities_Em2}. From the velocity map, we extract two types of information: the limit $v_y = 0$ between the attraction and the repulsion region (Fig. \ref{Fig_Velocities_Em2}(a)), and the velocity graphs at $\dy_{av}$ = constant (Fig. \ref{Fig_Velocities_Em2}(b)). The same procedure has been performed for all the emulsions and oils, and the five fluids can be compared in Fig. \ref{Fig_Contour_all}.

The shape of the velocity maps is the same for all fluids (Fig. \ref{Fig_Contour_all}(a)). Note that we do not observe an effect the fluid rheological properties on the location of the repulsion and attraction regions, within the accuracy of the experiments. When the bubbles are almost vertically aligned, \ie at lower values of $\dx_{av}$, they repel each other. 
For all the fluids, the contour $v_y = 0$ is located between $\dx_{av}/d = 1$ and $\dx_{av}/d = 2$. It seems that the repulsion region widens when the vertical distance between the bubbles increases, in other words, the $v_y = 0$ contour seems to make an angle $\theta \simeq 60 \degree$ with the flow direction. 
These velocity maps recall the hydrodynamic interaction between two rising bubbles in Newtonian fluid \cite{2011_Velez} where the trail bubble is attracted by the leading bubble if the angle to the flow direction (vertical) is less than 50 \degree, and is repelled otherwise. The experimental and theoretical studies have shown that side-by-side bubble repulsion can occur even at Reynolds numbers down to 0.02 \cite{2003_Legendre}, and in-line attraction has been observed at $\Rey \sim 0.1$ \cite{2011_Velez}.

In the example of Em2 in Fig.\ \ref{Fig_Velocities_Em2}(b), we see that at a given height difference $\dy_{av}$, the velocity profiles show a transition between the repulsion region ($v_y >0$) and the attraction region ($v_y < 0$). The transition is sharp at $\dy_{av}/d = 1 $ and become more smooth when $\dy_{av}$ increases. For $ 2 \lesssim \dx_{av}/d \lesssim 4$, the velocity curves exhibit a quasi-plateau for $\dy_{av}/d =$ 1 and 2.
Let us now compare the velocities measured in the different fluids. We observe in Fig. \ref{Fig_Contour_all}(b) that the plateau value in the attraction region is similar in all three emulsions and in V10K, even if the elastic parameter S is one order of magnitude lower in V10K than in emulsions. This shows that the elastic stresses $N_1$ do not affect the migration velocity. On the other hand, the absolute values of the velocity $v_y$ are one order of magnitude smaller in V100K than in the other fluids. At the studied shear rate ($\dot{\gamma} = 5~s^{-1}$), the viscosity of V100K is one order of magnitude larger than in the other fluids. The viscosity is therefore the major rheological property governing the cross-stream migration velocity, whereas we cannot observe any effect of the elasticity.

\subsubsection{Effect of geometry}
Lastly, we have investigated how robust our observations are to the geometry of the flow. Shear flow in a Couette cell is often used in rheometry experiments as a quasi-perfect uniform shear flow. However, the shear stress depends on the position in the gap. If $r$ is the distance to the rotation axis, angular momentum balance gives $\tau_{Couette} = T / (2 \pi r L)$, with $T$ the applied torque and $L$ the length of the rotating cylinder \cite{2008_Ovarlez}. With the Couette geometry dimensions that we are using, the stress at the inner wall is 9\% larger than at the external wall. The curvature of the Couette cell is known to affect the migration of suspended objects \cite{1979_Chan, 1981_Chan}.  In practice, for bubbles, which are deformable objects, we expect that the equilibrium position inside the gap is closer to the external wall than to the rotating cylinder \cite{1979_Chan, 2014_Villone}.

To check whether this affects the bubble-pair interaction, we have performed experiments with silicone oil V10K in a plate-plate geometry. Contrarily to the Couette cell, the shear rate is constant across the gap at a given location $r$. In addition, the plate-plate cell avoids buoyant bubble rising. However, between the plates, the shear rate depends on the distance to the rotation axis $r$: $\dot{\gamma}_{Plates} = \Omega r / h$, where $\Omega$ is the rotational velocity of the upper plate. Therefore, bubbles located at different distances from the center experience different shear rates.

\begin{figure}[t]
\begin{center}
\includegraphics[height=5cm]{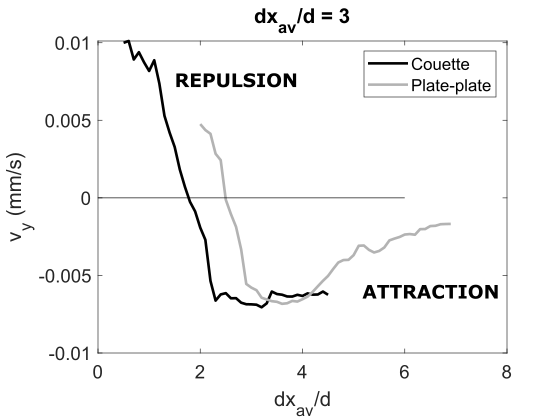}
\end{center}
\caption{Relative velocity in the direction of the vorticity in V10K, in the Couette cell and in the plate-plate geometry (h = 1.5 mm), at $\dy_{av}$/d = 3.}
\label{Fig_Geometries}
\end{figure}

In Fig. \ref{Fig_Geometries}, we can first compare the velocity profiles obtained with both geometries for the same gap size, $h = 1.5$ mm. Between the plates like in the Couette cell, the bubbles repel each other when they are aligned in the vorticity direction, and attract each other otherwise. The measured velocities are similar in both geometries in the attraction region. The geometry-dependent gradients of the shear rate do not therefore affect noticeably the bubble-pair interaction, and the observations are robust on the geometry of the flow.

\subsection{Bubble-pair simulations}\label{ssec:pair_sims}

In this section, the results of bubble-pair simulations are presented, to further explain experimental observations. Simulations have been conducted with the bubble pairs in various initial configurations, some where the bubbles are aligned in the vorticity direction, and some where they are not (\ie $\mathrm{d}x \neq 0$). Simulations have been conducted for emulsion and silicone oil cases; see Table \ref{tab:params} for the dimensionless numbers used in the simulations. A summary of the different types of bubble-pair interactions observed experimentally and by numerical simulation, in both oils and emulsions, is presented in Fig. \ref{fig:summary_2bubbles}. Below, we first discuss the simulation results.

\subsubsection{Attractive cases}\label{sssec:numeric_attr}
\begin{figure}[!ht]
 \centering
 \begin{subfigure}[b]{0.385\textwidth}
 \centering
 \caption{}
 \includegraphics[width=\textwidth]{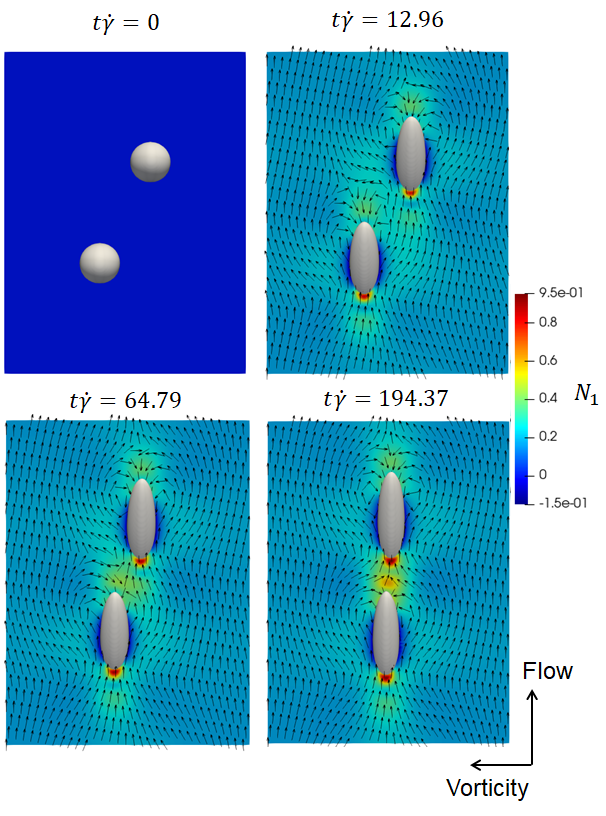}
 \label{fig:oil_attr_intfEvna}
 \end{subfigure}
 \hspace{0.1cm}
 \begin{subfigure}[b]{0.35\textwidth}
 \centering
  \caption{}
 \includegraphics[width=0.9\textwidth]{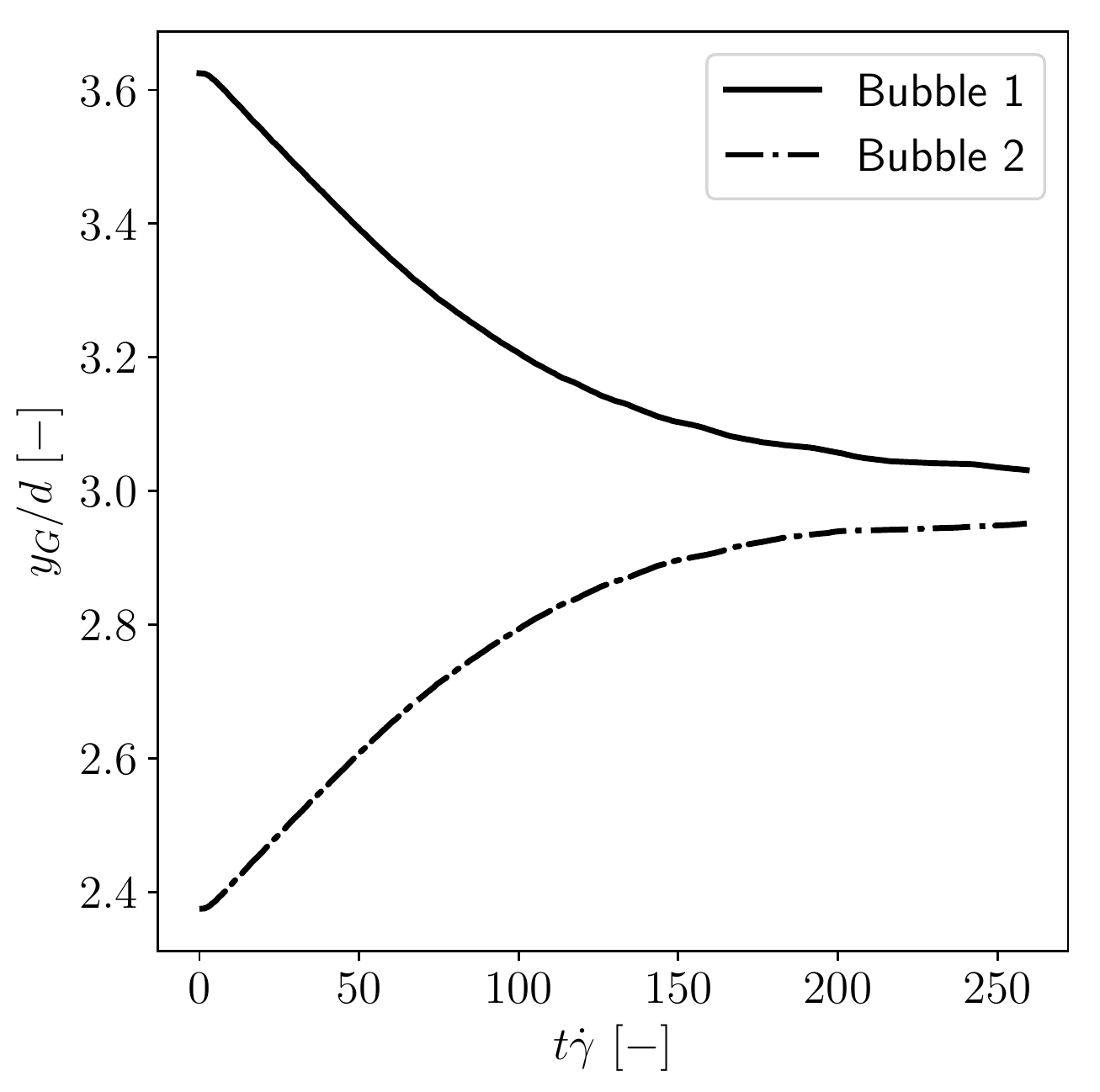}
 \label{fig:oil_attr_traj}
 \end{subfigure}
 \caption{Motion of bubble-pair suspended in a weakly VE silicone oil when they are initially not aligned in the vorticity direction. (a) Visualisation of the evolution of the bubble-oil interface at four snapshots in time, as well as the $N_1$ contours and streamlines on the central XY plane. (b) COM positions in the vorticity direction of the bubbles as a function of time. Bubble 1 corresponds to the right bubble in each snapshot, and Bubble 2 the left. $\Rey = 0.05$, $\Ca = 0.5$, $\Wi = 0.1$, initial separation distance of the COM of the bubbles: $\dx_0=2.5d$, $\dy_0=1.25d$.}
\label{fig:oil_attr_intfEvn}
\end{figure}

For the simulations corresponding to case of silicone oils, when $\dx_0 = 2.5d$\, bubble alignment under shear flow of the bubbles is observed \ie the bubble pair attracts in the vorticity direction. In \figref{fig:oil_attr_intfEvn}, the bubble shapes are visualised at four different instances in time. It can be seen that when $\dx$ is sufficiently high, alignment under shear flow is observed, which successfully captures the phenomenology of the experimental observation. Furthermore, the trajectories of the centers of mass (COM) in the vorticity direction, $y_G$, of the two bubbles are shown in Fig. \ref{fig:oil_attr_traj}; the bubbles are indexed according to Fig. \ref{fig:oil_attr_intfEvna} \ie Bubble 1 corresponds to the right bubble, and Bubble 2 to the left. The time evolution of $y_G$ in Fig.\ \ref{fig:oil_attr_traj} shows that the bubbles approach each other in the vorticity direction and are almost aligned after the simulation has run for $\approx 250$ time units. 

\begin{figure}
 \centering
\begin{subfigure}[b]{0.39\textwidth}
\centering
 \caption{}
\includegraphics[width=\textwidth]{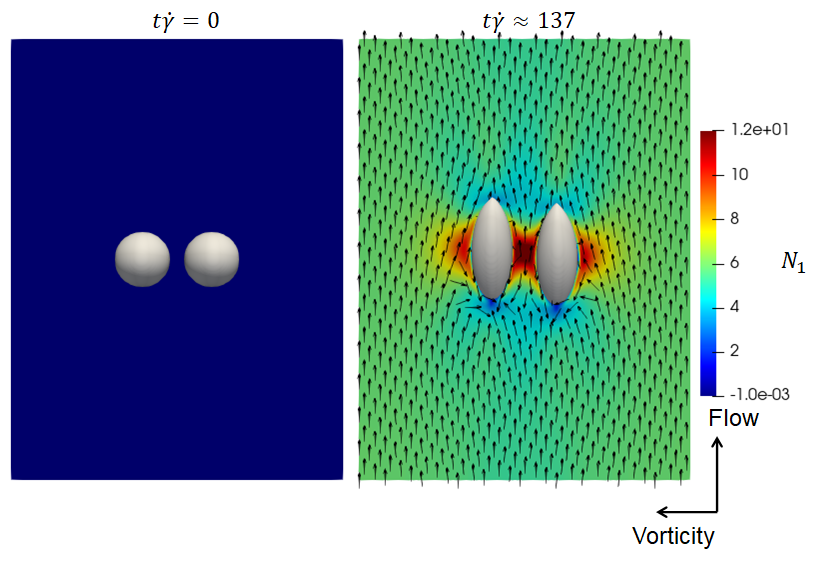}
\label{fig:Em3_rep_intfEvna}
\end{subfigure}
\hspace{0.1cm}
\begin{subfigure}[b]{0.35\textwidth}
\centering
 \caption{}
 \includegraphics[width=\textwidth]{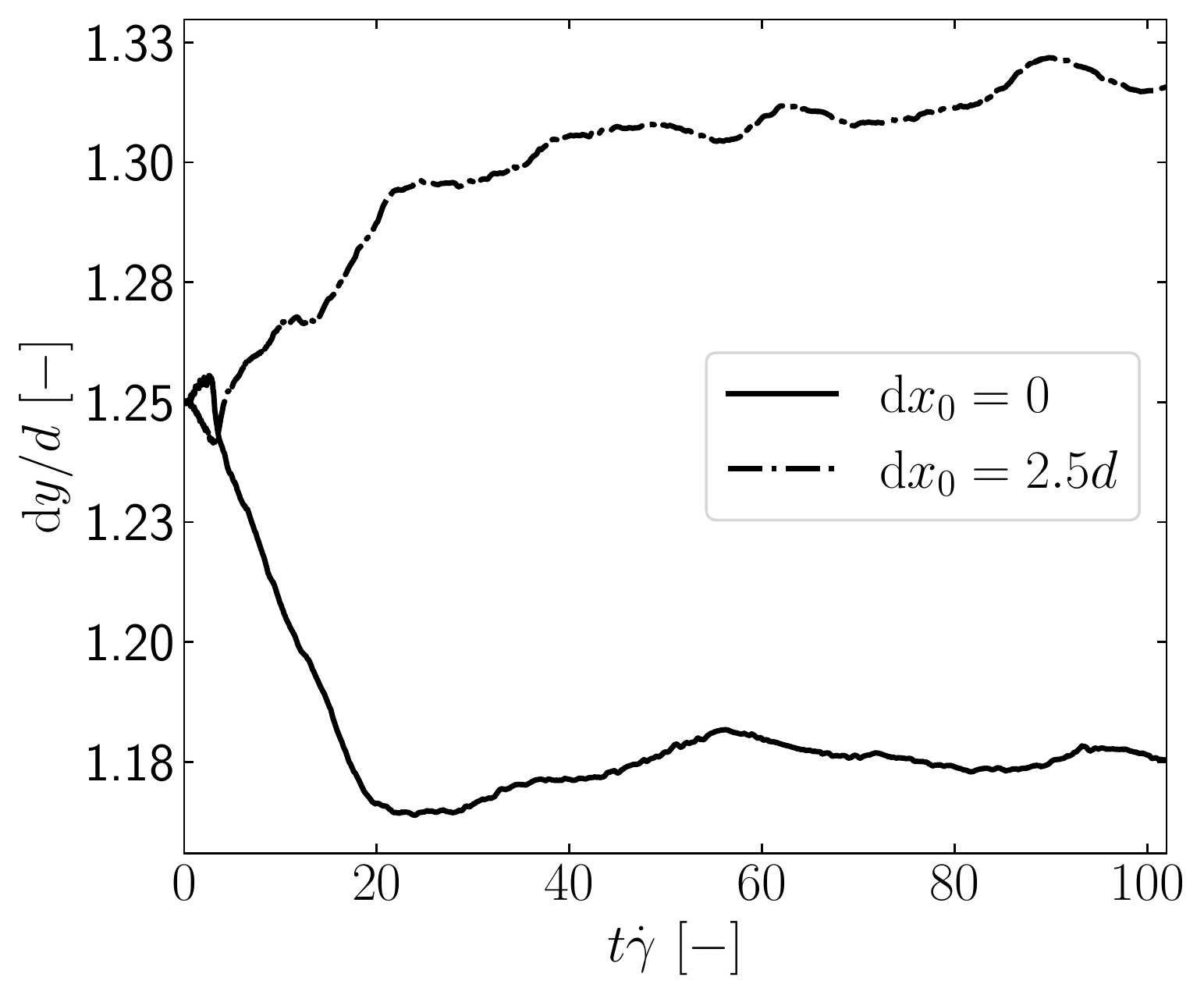}
\label{fig:Em3_rep_traj}
\end{subfigure}
 \caption{ Average relative position in the vorticity direction as a function of time for two $\dx_0$ in Emulsion Em3, simulated with the EVP model with $\Rey = 0.05$, $\Ca = 0.5$, $\Wi = 7.023$, $\alpha_s = 1./9$, $\Bing = 0.422$, $n=0.5$, $\dy_0=1.25d$.}
\label{fig:Em3_rep_intfEvn_traj}
\end{figure}

Emulsions have been simulated with the Saramito Herschel-Bulkley EVP model. Two parameter fitting methods were used for setting $\Wi$ for the numerical simulations of the emulsion cases. In the first method, $\Wi$ was set so that the simulations produce the experimentally measured $S$. Setting $\Wi$ like this produces a range of $\Wi$ of $3 \leq Wi \leq 9$ for the three emulsions. Bingham numbers up to $0.5$ were studied. In these simulations, in the configurations where attraction is experimentally observed (\ie $\dx_0 = 2.5d$), very little bubble migration is observed. As shown in \figref{fig:Em3_rep_intfEvn_traj}, the distance $\dy$ reaches a plateau value after 20 time units 5\% above its initial value. We will discuss in \secref{sec:numeric_discussion} the possible reasons for the absence of migration in the simulated emulsions in this case.

\begin{figure}
\includegraphics[width=0.45\textwidth]{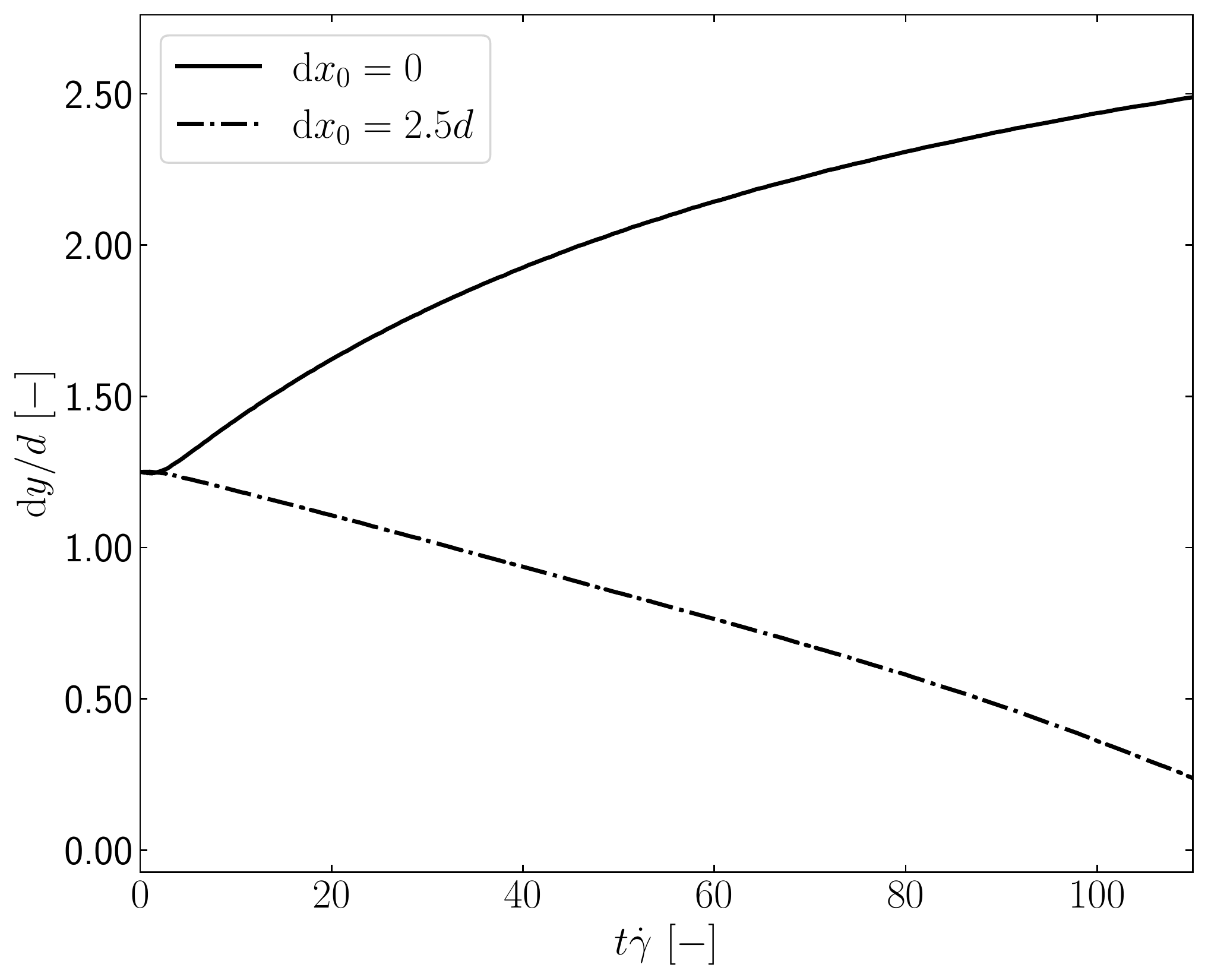}
\caption{Average relative position in the vorticity direction as a function of time for two $\dx_0$ in Emulsion Em3 with the EVP model. $\Rey = 0.05$, $\Ca = 0.34$, $\Wi=0.066$, $\alpha_s = 1./9$, $\Bing = 0.422$, $n=0.5$.}
\label{fig:Em3_fit2_dytraj}
\end{figure}

A second method for fitting the model parameters and setting $\Wi$ was also used for the simulations. As before, $k$, $n$, and $\tau_y$ are set according to the experimental fitting of the viscoplastic Herschel-Bulkley model. The elastic modulus, $G$, however, is set as the low strain amplitude $\Gp$ which was measured in oscillatory shear tests of the emulsions, following \cite{prudhomme1987response, Tsambubble}. This elastic modulus $G$ is then used to get a characteristic relaxation time, $\lambda$, for the flow in order to set $\Wi$, giving $\Wi = 0.066$ for emulsion Em3. The relative bubble-pair trajectory in emulsion Em3 is plotted as a function of time in \figref{fig:Em3_fit2_dytraj}. Here, it can be seen that the relative bubble-pair trajectory shows much better agreement with the phenomenological observations of the experiments: when the bubble-pair is aligned in the vorticity direction ($\dx_0 = 0$), they repel each other and $\dy$ increases with time; when the bubble-pair becomes sufficiently separated in the streamwise direction ($\dx_0 = 2.5d$), they attract each other and $\dy$ decreases with time. The difference in the simulation results for the two parameter fittings are discussed in \secref{sec:numeric_discussion}.

\subsubsection{Repulsive cases}\label{sssec:numeric_rep}

\begin{figure}
 \centering
 \begin{subfigure}[b]{0.39\textwidth}
 \centering
  \caption{}
 \includegraphics[width=\textwidth]{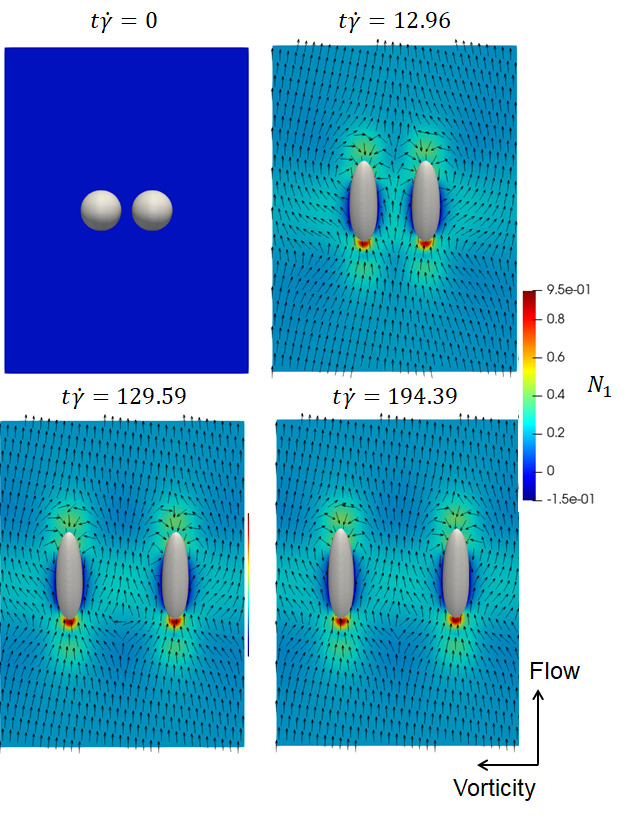}
 \label{fig:oil_rep_intfEvna}
 \end{subfigure}
 \hspace{0.1cm}
 \begin{subfigure}[b]{0.35\textwidth}
 \centering
  \caption{}
 \includegraphics[width=0.9\textwidth]{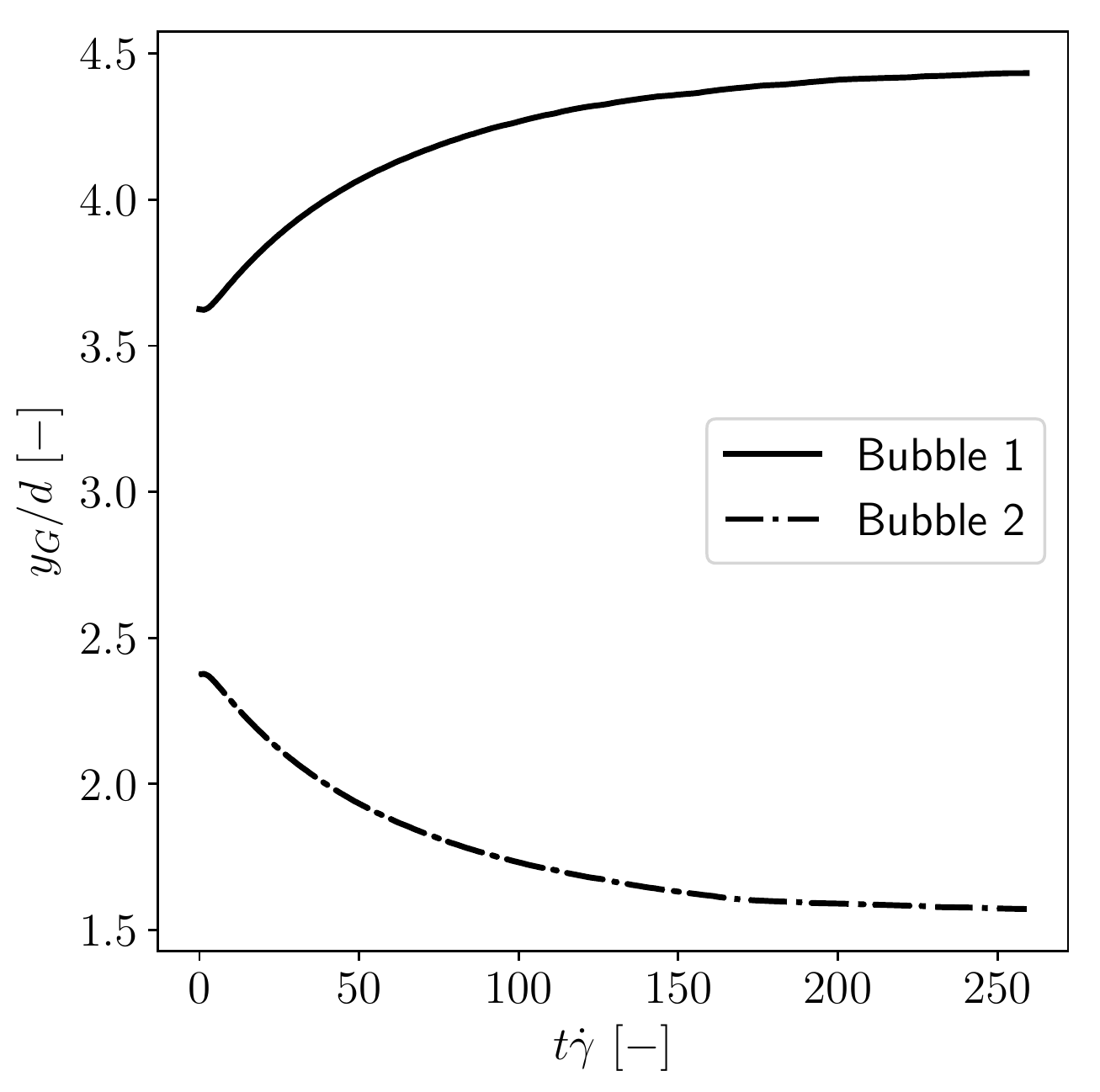}
 \label{fig:oil_rep_traj}
 \end{subfigure}
 \caption{Motion of bubble-pair suspended in a weakly VE silicone oil when they are initially aligned in the vorticity direction. (a) Visualisation of the evolution of the bubble-oil interface at four snapshots in time, as well as the $N_1$ contours and streamlines on the central XY plane. (b) COM positions of bubbles 1 and 2 in the vorticity direction as a function of time. Bubble 1 corresponds to the right bubble in each snapshot, and Bubble 2 the left. $\Rey = 0.05$, $\Ca = 0.5$, $\Wi = 0.1$, initial separation distance of the COM of the bubbles: $\dx_0 = 0$, $\dy_0=1.25d$.}
\label{fig:oil_rep_intfEvn}
\end{figure}

In all bubble-pair experiments, both in emulsions and silicone oils, repulsion was observed when the bubble pair was initially almost aligned in the vorticity direction. We present the corresponding simulations of a bubble pair in a silicone oil suspending phase in Fig.\ \ref{fig:oil_rep_intfEvna}, where the evolution of the bubble-oil interface is visualised at four different instants, together with the contours of the first normal-stress difference $N_1$ and  the streamlines on the central XY plane. 
Indeed, when the bubble pair is initially aligned in the vorticity direction, repulsion is observed in the simulation. As in section \ref{sssec:numeric_attr}, the coordinate $y_G$ of each bubble is shown as a function of time in Fig.\ \ref{fig:oil_rep_traj}; it can be seen that $y_G$ evolves such that bubbles 1 and 2 repel each other in the vorticity direction; after the simulation has run for $\approx 250$ time units, $y_G$ reaches a steady state. This is because the two bubbles get sufficiently far apart that the repulsive force between them diminishes, and they experience the periodic boundary effects.

For simulated emulsions, in the repulsive configuration - like in the attractive configuration - very little bubble migration is observed in the case Wi = 7 where the Weissenberg number is deduced from the experimental elastic number S. An example of the trajectory is shown in \figref{fig:Em3_rep_intfEvn_traj} with $\dx_0 = 0$. With the second fitting method giving Wi = 0.066, repulsion is observed in the simulated trajectories (see \figref{fig:Em3_fit2_dytraj}). These observations will be addressed in the results discussion in \secref{ssec:num_discp}.

\subsubsection{Migration velocities}\label{ssec:migvel}
\begin{figure}
\includegraphics[width=0.45\textwidth]{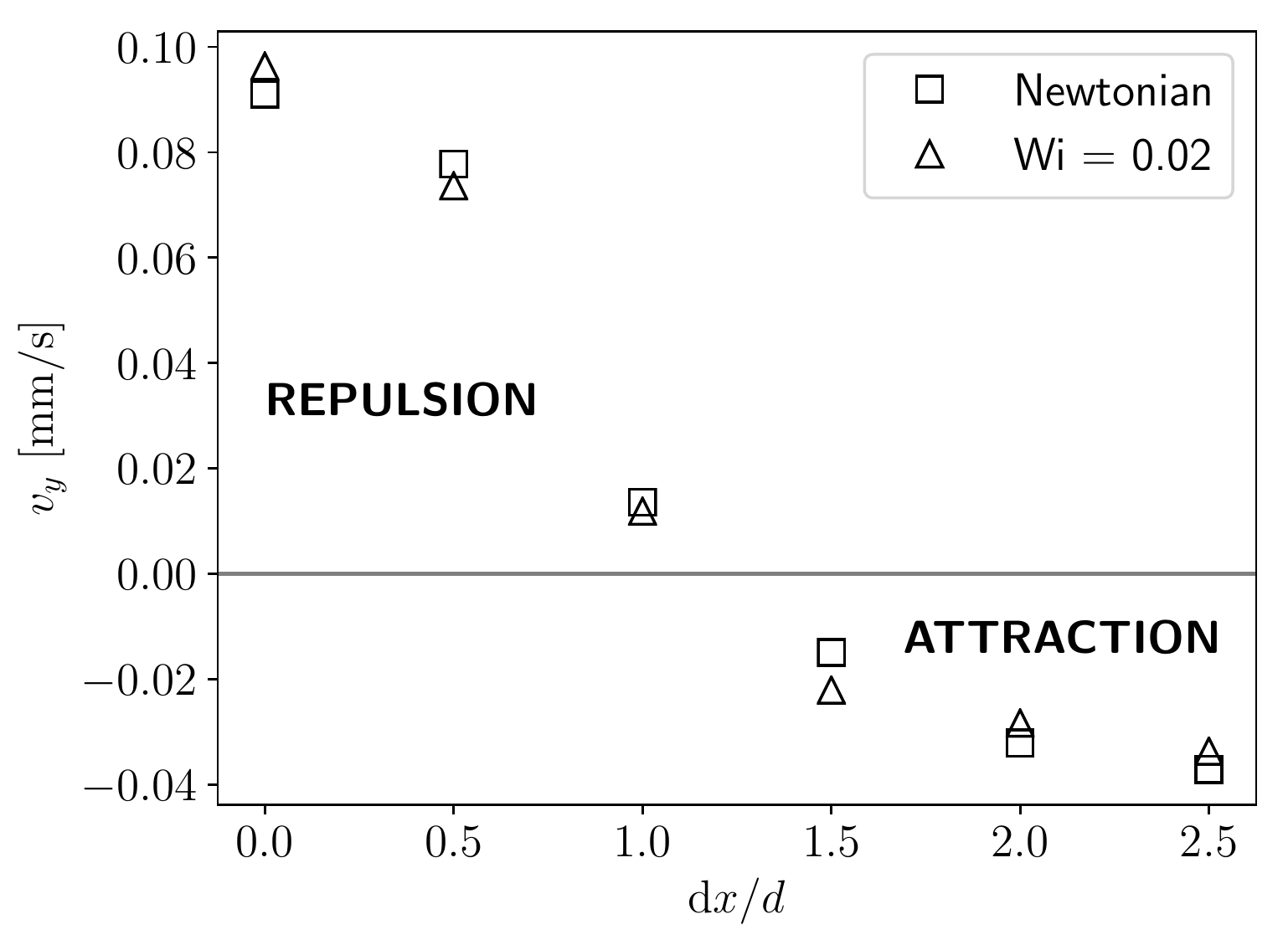}
\caption{Migration velocity of the bubble-pair in Newtonian suspending phase, and weakly VE silicone oil suspending phase as a function of the bubble separation distance in the flow direction. $\Rey = 0.05$, $\Ca = 0.5$, $\mathrm{d}y/d = 2$.}
\label{fig:vel_transition_numeric}
\end{figure}

\begin{figure}
\includegraphics[width=0.45\textwidth]{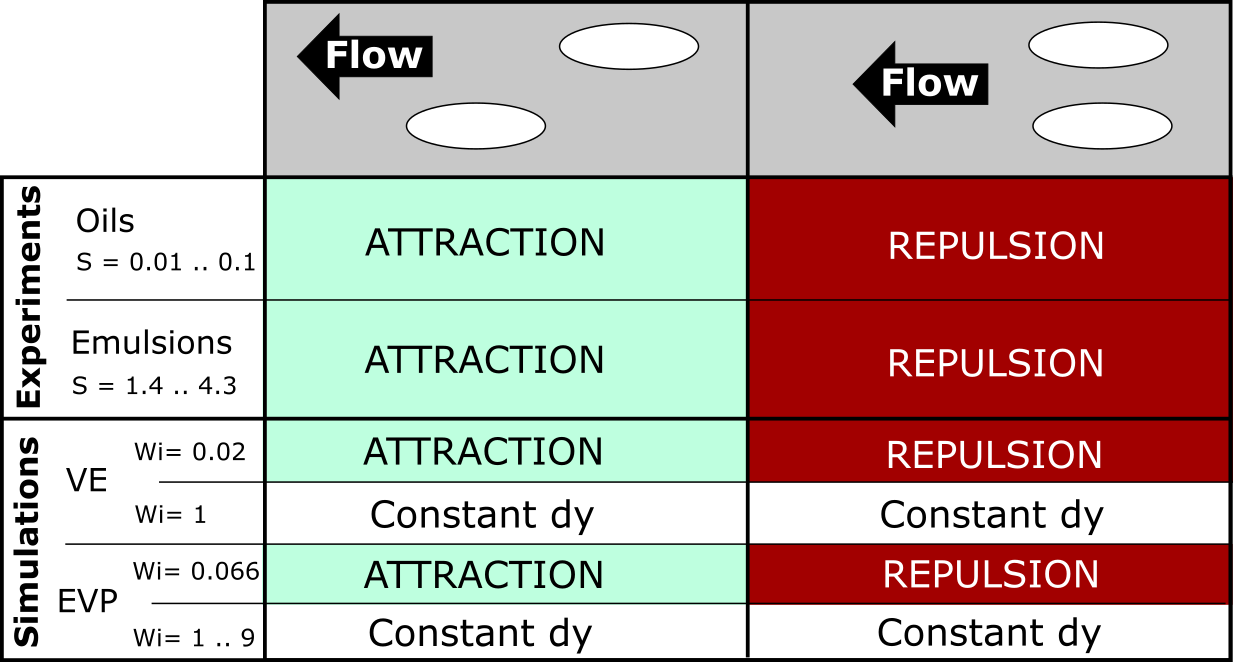}
\centering
\caption{Summary of the type of interactions observed in the experiments and simulations, in oils and emulsions.}
\label{fig:summary_2bubbles}
\end{figure}

It was observed experimentally for both oils and emulsions, and numerically for the case of oils, that two different migration patterns occur depending on the distance between the bubbles in the flow direction. 
As for the experimental data in Fig.\ \ref{Fig_Contour_all}, a plot of the bubble migration velocity as a function of initial distance between their COM in the flow direction is presented in Fig. \ref{fig:vel_transition_numeric} for the simulations modelling the dynamics in oil. A similar transition from repulsion (positive $v_y$) to attraction (negative $v_y$) with increasing $\mathrm{d}x$ is observed in the numerical data - $v_y$ is calculated to be the same order of magnitude as in experiments, although the migration velocities are somewhat over-predicted by the Oldroyd-B model. Preliminary results have shown that $v_y$ decreases with increasing $k_\mu$, so the over-prediction in $v_y$ may be due to the lower value of $k_\mu$ used in the simulations compared to the experiments.

\section{Discussion}\label{sec:numeric_discussion}

\subsection{The role of bubble deformability}\label{ssec:deformability}

It is well-known that a rigid sphere under creeping flow conditions in a Newtonian continuous phase does not show any lateral migration with respect to the primary flow direction (as this would violate the reversibility of the system). The object deformability on the other hand breaks the reversibility  as it adds non-linearity  to the system and lateral migration of a particle, droplet or bubble in the continuous phase can take place, as found in numerous works \eg a neo-Hookean particle in a Newtonian continuous phase under confined simple shear flow migrates to the channel center \cite{2014_Villone,Dhiya19}; a viscous drop in a Newtonian continuous phase placed near a wall migrates away from it under simple shear \cite{goldsmith1963flow, mukherjee2013effects}.

\begin{figure}[tb]
 \centering
 \begin{subfigure}[b]{0.35\textwidth}
 \centering
  \caption{}
 \includegraphics[width=0.9\textwidth]{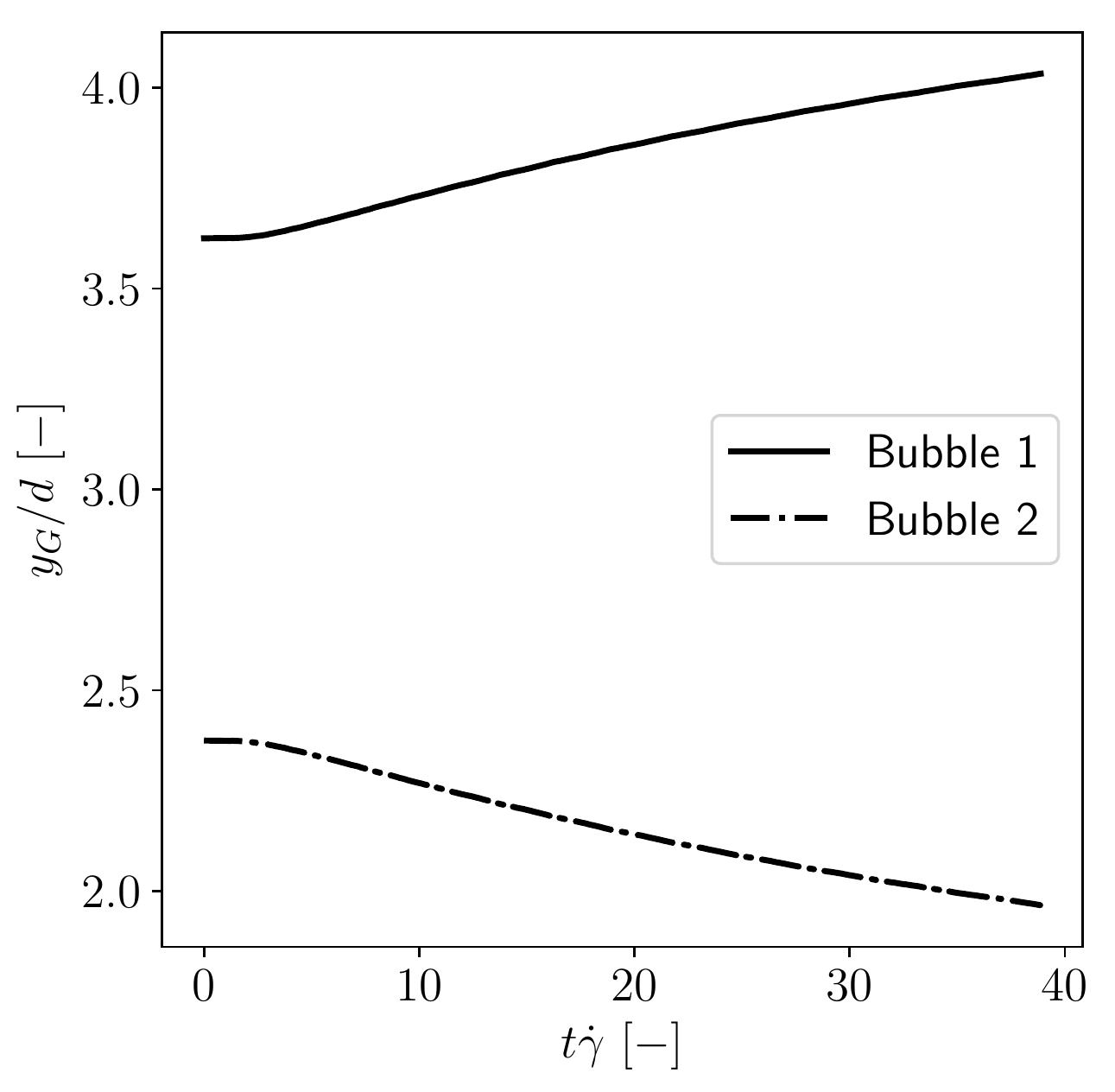}
 \label{fig:deform_bubs}
 \end{subfigure}
 \hspace{0.1cm}
 \begin{subfigure}[b]{0.35\textwidth}
 \centering
 \caption{}
 \includegraphics[width=0.9\textwidth]{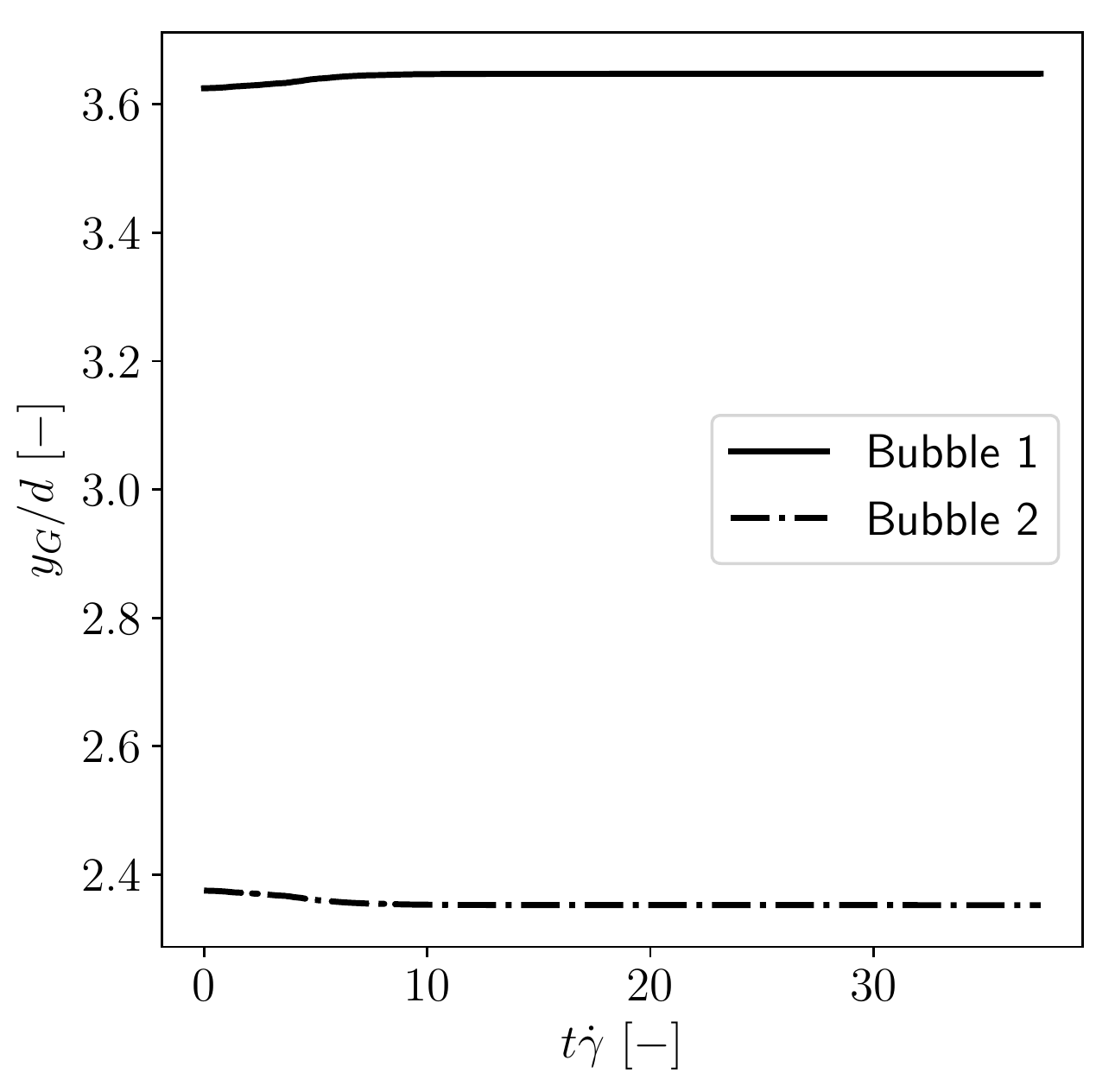}
 \label{fig:stiff_bubs}
 \end{subfigure}
 \caption{Trajectory of the COM position in the vorticity direction for a bubble-pair initially aligned in the vorticity direction in a Newtonian suspending phase for $\Rey = 0.05$ and  (a) $\Ca = 0.5$, (b) $\Ca = 0.05$. The initial configuration of the bubbles, and their indexing (\ie Bubble 1 and 2), are as in \figref{fig:oil_rep_intfEvn}.}
\label{fig:deform_vs_stiff}
\end{figure}

In sections \ref{ssec:pair_exps} and \ref{ssec:pair_sims}, we have presented results on interactions between bubble pairs and the lateral migration in the vorticity direction that was observed in both simulations and experiments. We have numerically studied bubble pairs under the same conditions as in section \ref{ssec:pair_sims} with a Newtonian suspending phase and have observed the same lateral migration patterns in the vorticity direction \ie repulsion when initially aligned in the vorticity direction, a decrease in $v_y$ with increasing separation in the flow direction, and a repulsion-to-attraction transition when they are separated sufficiently far in the flow direction. This is shown in Fig. \ref{fig:vel_transition_numeric} where the same repulsion-attraction transition is seen for a Newtonian case as well as for a weakly VE ($\Wi = 0.02$) case. These simulation results, alongside the same experimental observation in the weakly VE V10K oil ($S \sim 0.01$), lead us to hypothesise that bubble deformability also plays a role in these lateral migration patterns that were observed. Furthermore, we have simulated a case with stiffer bubbles at a much lower Capillary number ($\Ca = 0.05$) for comparison with the $\Ca = 0.5$ case, presented in Fig. \ref{fig:deform_vs_stiff}. It can be seen in the $y_G$ trajectory that while in the $\Ca = 0.5$ case, the bubbles repel each other, in the $\Ca = 0.05$ case, the bubbles only repel each other initially (possibly due to some deformation effects still present), after which they do not migrate. Similarly, no migration is observed in an attraction case if $\Ca = 0.05$. Thus, we conclude that bubble deformability in a \textit{Newtonian} suspending phase can cause this same observed repulsive-attractive lateral migration in the vorticity direction for bubble pairs. As such, the bubble deformability may play a similar role in a VE/EVP suspending phase, in addition to the elastic effects which then become introduced.

\begin{figure*}[ht]
 \centering
 \begin{subfigure}[b]{\parrw\textwidth}
 \centering
 \caption{}
 \includegraphics[height=4.5cm]{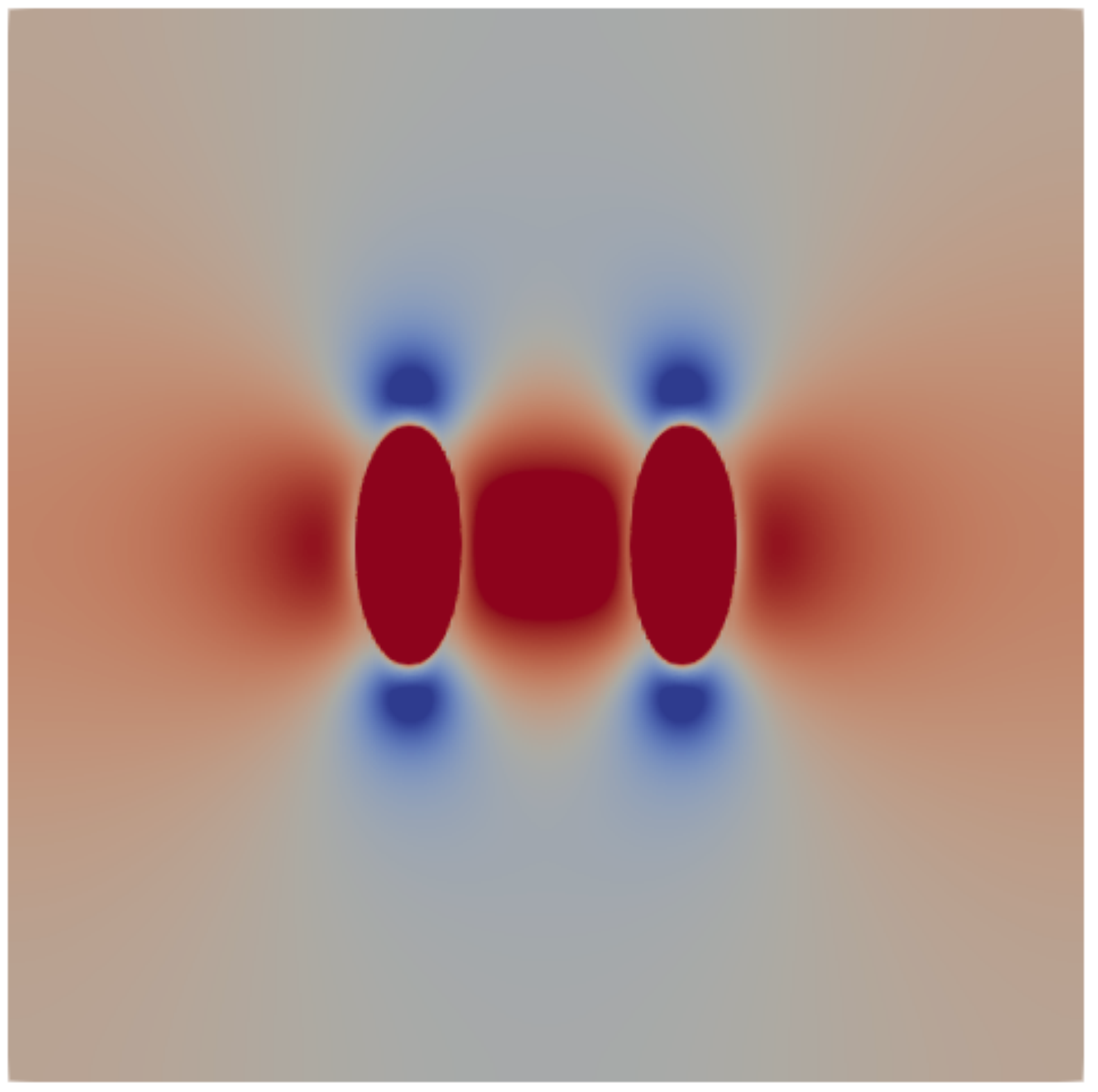}
 \label{fig:p_dx0p0}
 \end{subfigure}
 \begin{subfigure}[b]{\parrw\textwidth}
 \centering
  \caption{}
 \includegraphics[height=4.5cm]{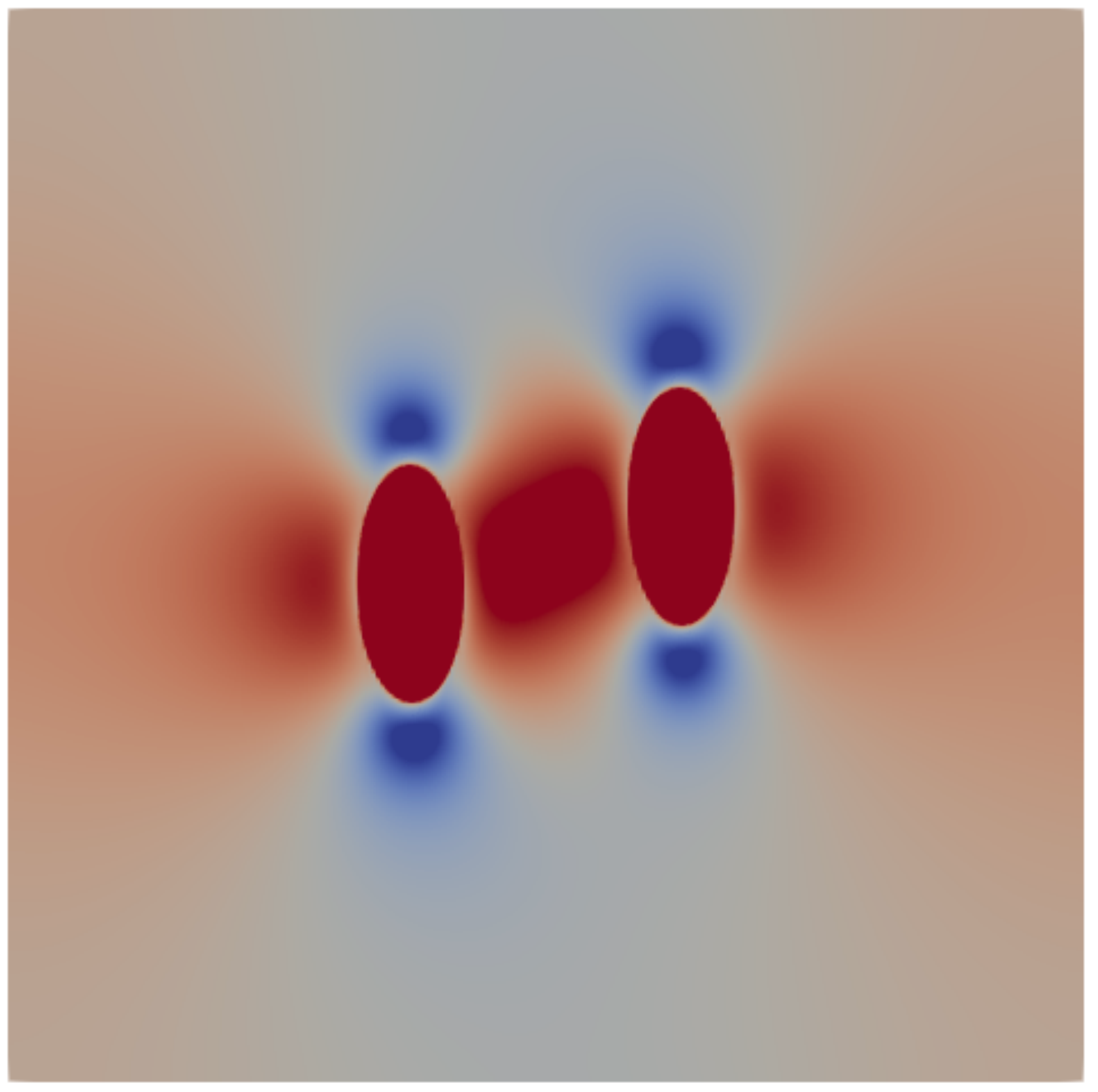}
 \label{fig:p_dx0p5}
 \end{subfigure}
 \begin{subfigure}[b]{\parrw\textwidth}
 \centering
  \caption{}
 \includegraphics[height=4.5cm]{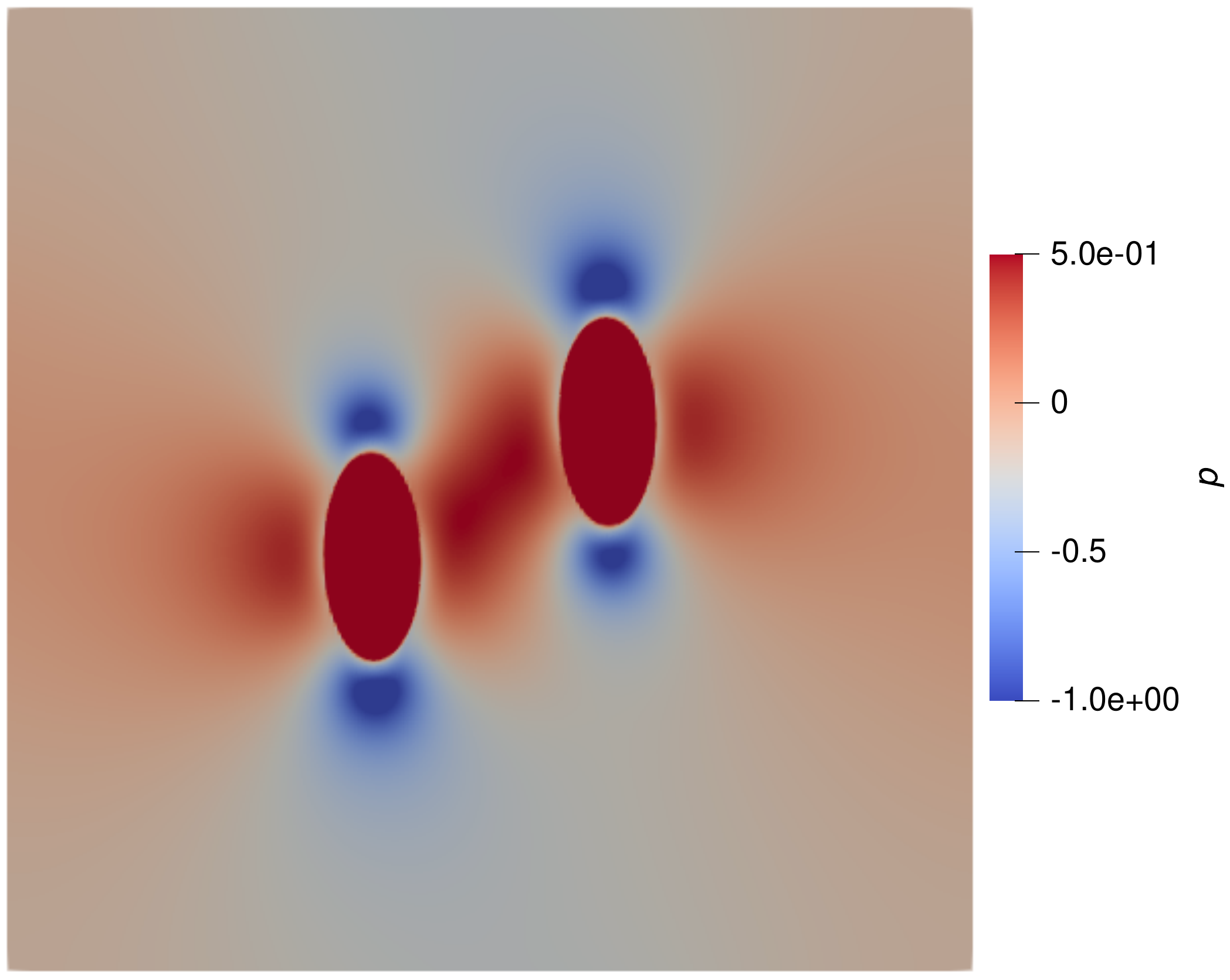}
 \label{fig:p_dx1p0}
 \end{subfigure}
 
 \begin{subfigure}[b]{\parrw\textwidth}
 \centering
  \caption{}
 \includegraphics[height=4.5cm]{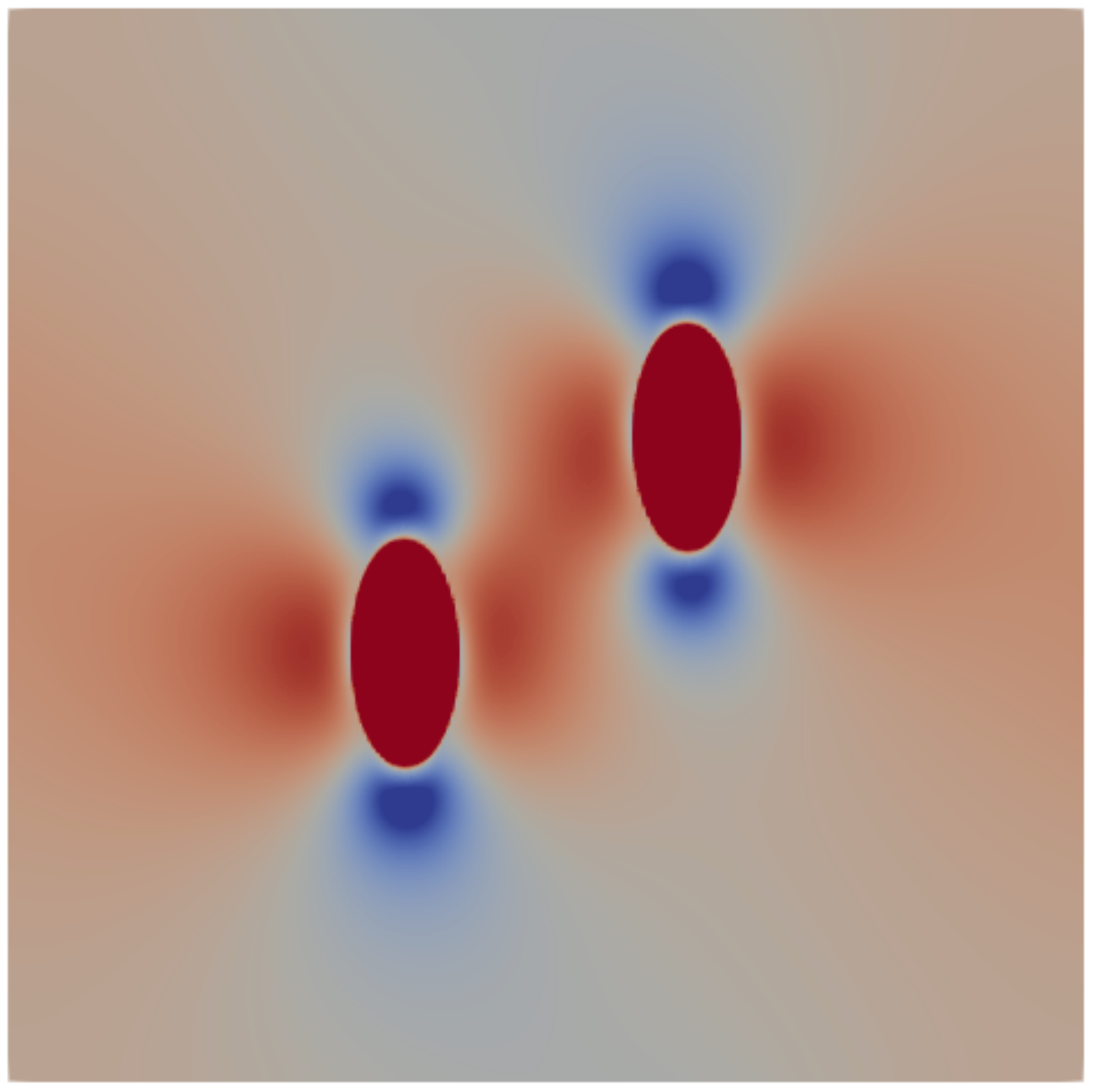}
 \label{fig:p_dx1p5}
 \end{subfigure}
 \begin{subfigure}[b]{\parrw\textwidth}
 \centering
  \caption{}
 \includegraphics[height=4.5cm]{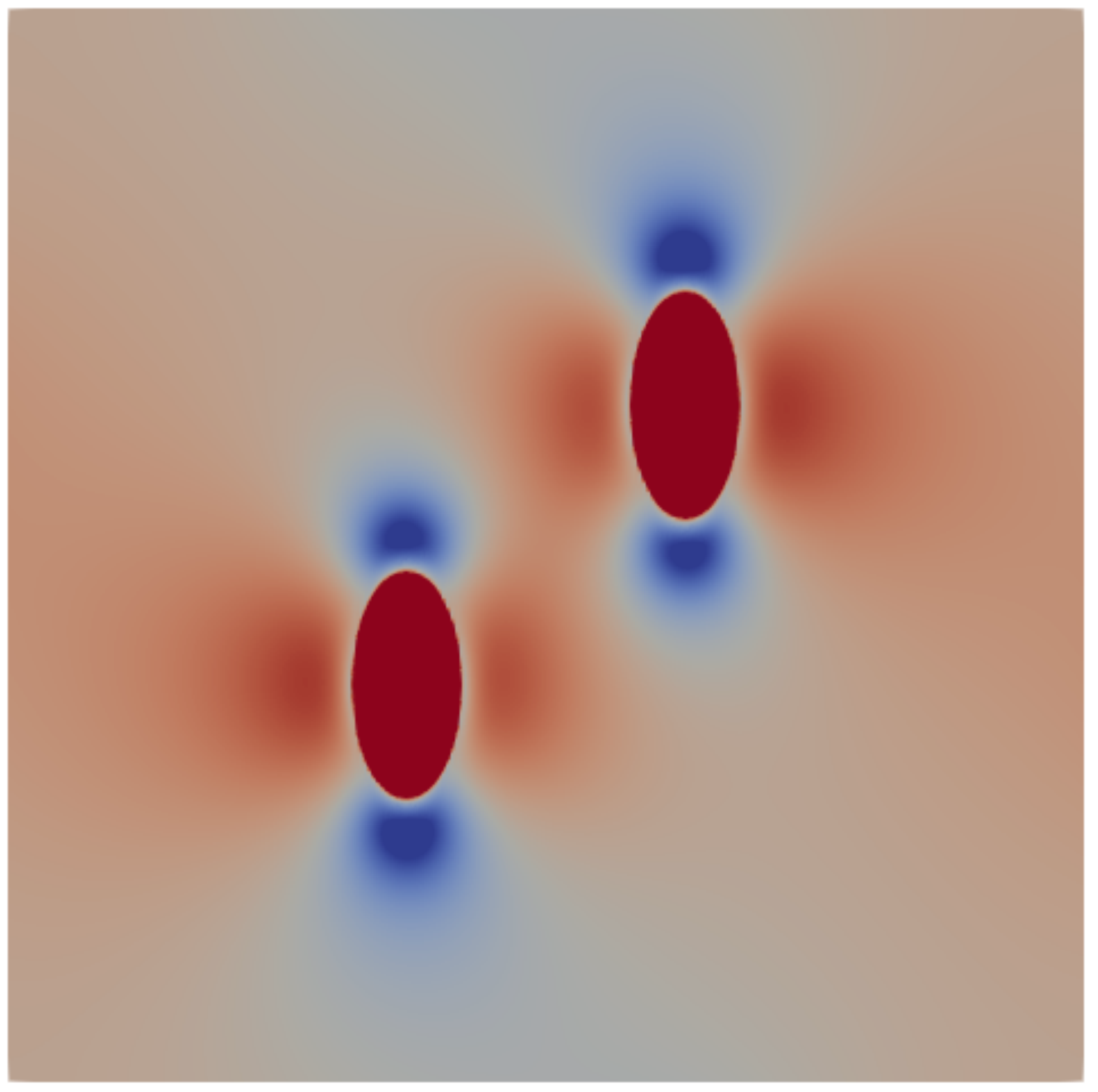}
 \label{fig:p_dx2p0}
 \end{subfigure}
 \begin{subfigure}[b]{\parrw\textwidth}
 \centering
  \caption{}
 \includegraphics[height=4.5cm]{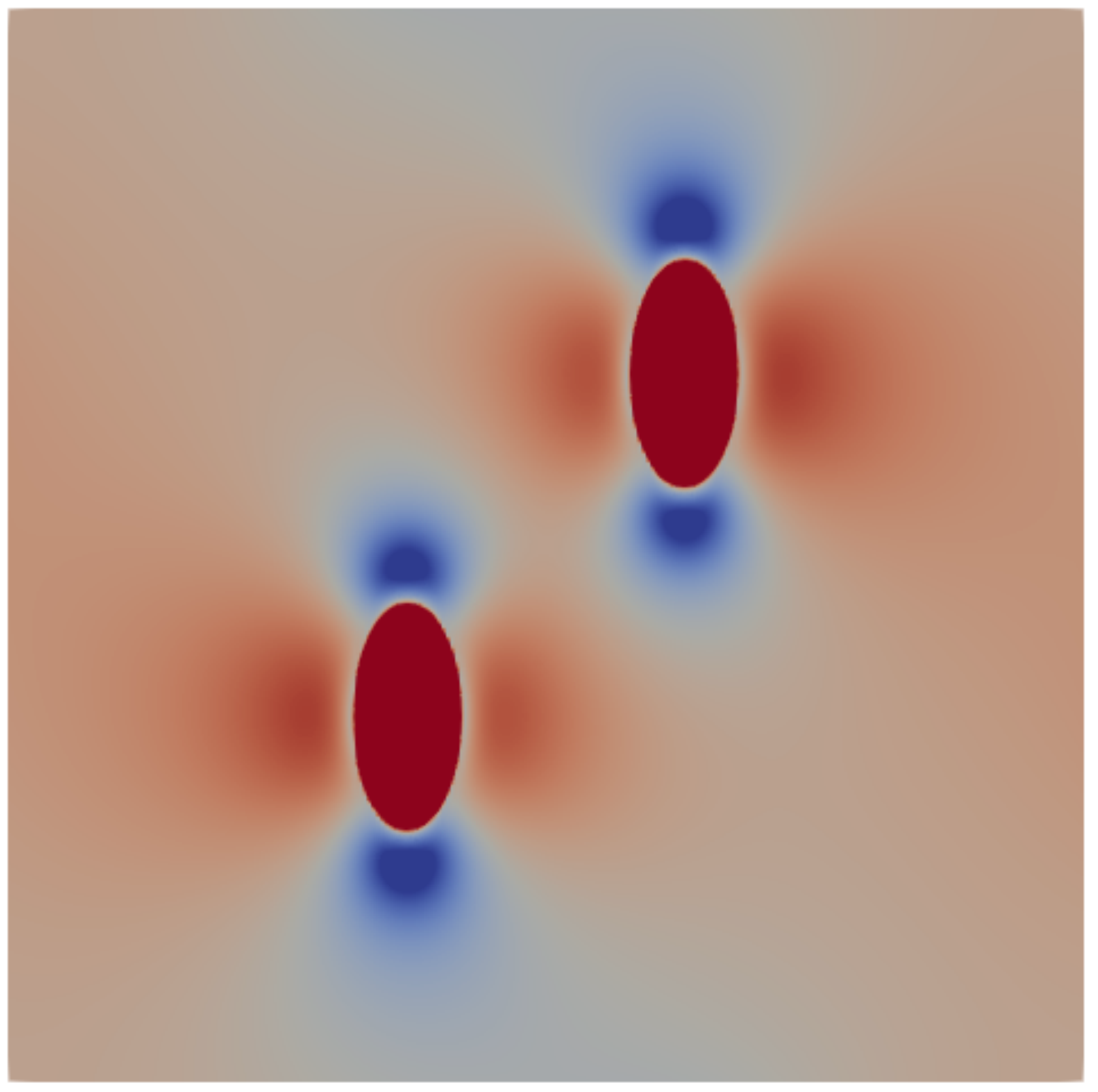}
 \label{fig:p_dx2p5}
 \end{subfigure}
 
 \caption{Pressure fields on the XY plane for bubble-pairs at different initial positions: (a) $\dx_0 = 0$, (b) $\dx_0 = 0.5d$, (c) $\dx_0 = d$, (d) $\dx_0 = 1.5d$, (e) $\dx_0 = 2d$, (f) $\dx_0 = 2.5d$. The bubbles are suspended in the viscoelastic silicone oil \ie $\Rey = 0.05$, $\Ca = 0.5$, $\Wi = 0.02$, $\alpha_s = 1/9$. The cases are all visualized at the same dimensionless time, $t\dot{\gamma}\approx 9.72$ time units.}
\label{fig:pos_sweep_pressure}
\end{figure*}

To explain how the deformability of the bubble-pair affects the flow dynamics to cause the migration, we have analyzed the pressure field around the bubble-pair to identify the mechanism that is responsible for the migration. In \figref{fig:pos_sweep_pressure}, we display the pressure contours on the XY plane of the bubble-pair in VE silicone oil for different $\dx_0$, at the same dimensionless time. It can clearly be seen that there is a large pressure imbalance on the bubbles in the vorticity direction. \figref{fig:p_dx0p0} shows that the region of fluid confined between the two bubbles has a higher pressure than the region on the "outer" side of the bubble. Examining the behavior of a single deformable bubble, we see that the pressure field around it changes such that large pressure regions form in the suspending fluid at the middle of the elongated side, and low pressure regions form near the tips of the bubble owing to the balance of pressure and surface tension at the fluid interface \cite{yue2005viscoelastic}. The interaction of these pressure fields leads to an unbalanced pressure distribution around each bubble. When the bubbles are initially aligned in the vorticity direction, a higher pressure region forms in the fluid region between the two bubbles, creating a thrust which causes them to repel each other. As $\dx_0$ increases, the interaction between the low pressure region near the tip and high pressure near the equator causes a pressure distribution such that the pressure is greater on the "outer" side of the bubble, leading to an attractive migration force and the bubbles align to minimise this pressure imbalance. This heterogeneous pressure distribution which forms around each bubble due to the capillarity and the interactions of the pressure fields of the bubble-pair is responsible for the migration.

Experiments on bubble pairs have been performed for Capillary numbers $\Ca \gtrsim 1$, \ie for deformable bubble, and interaction has been obtained in all the cases. Experiments on bubbles suspensions in emulsions, on the other hand, involve Capillary numbers down to 0.2. At the smallest values investigated, the final configuration appeared to be distant alignments, with inter-bubble distance being several times larger than their diameter. It is highly possible that this results from a reduced attraction for the less deformable bubbles, which supports our hypothesis that the deformation plays the main role in the bubble interaction studied here. Though, the presence of distant alignments indicate that the rheological properties of the emulsion, elasticity and shear thinning, also affect the bubble migration in the experiments.

\subsection{Differences between experiments and simulations}\label{ssec:num_discp}
We have observed a phenomenological agreement between experiments and simulations in the case of bubble pairs in a weakly elastic oil (VE model): attraction in the vorticity direction is observed when the bubble pair is not aligned in the vorticity direction, and a transition to repulsive migration occurs when the initial distance  in the flow direction, $\dx$, decreases. Experimentally, this behavior is very robust: it has been observed in oils of two different viscosities (i.e.\ two different values of the Reynolds and the capillary number) and in emulsions, and in two different shear geometries (Couette and plate-plate).

Simulations with the EVP model are expected to be representative of the experiments in emulsions. However, compared to the silicone oils, the emulsions are much more complex - shear-thinning behavior, strong elasticity ($S = 1 - 8$) and plasticity - all of which may cause the different observation in the simulation. We can therefore wonder how well the numerical models capture the complexity of the fluids. 
The EVP model with the Saramito-Herschel Bulkley equation has been selected because it includes viscous effects (quantified by the viscosities $\mu_{s,1}$ and $\mu_{e,1}$), plasticity (quantified by the yield stress $\tau_y$), elasticity (characterized by the relaxation time $\lambda$), and shear-thinning (characterized by the flow index $n$). 
These parameters have a physical relevance when they are applied to polymer suspensions: the viscosities correspond to the suspending phase and the polymers, and $\lambda$ refers to the relaxation of the polymer chains. The emulsions have a different microscopic structure than polymers: they contained highly packed oil droplets in a continuous aqueous phase. The model parameters have been chosen to fit the experimental rheological properties  from Fig. \ref{Fig_rheo}, but do not have a straightforward physical meaning. Furthermore, the model does not allow to fit all the experimental parameter simultaneously, \ie the viscosity curve, the first normal stress difference (or Elastic number S) and the elastic modulus, and we need to define which parameters must be prioritized. Two fitting methods have been investigated and compared, the first one prioritizes the first normal stress difference $N_1$ measured from the steady state curve, and the second one targets the elastic modulus $G'$ obtained from the oscillatory shear measurements in the linear viscoelastic regime. Simulations using the latter fitting yielded much better phenomenological agreement with the experiments, as seen in \figref{fig:Em3_fit2_dytraj} and seems therefore the most appropriate method for the configuration studied in this paper. However, it should be emphasised here that with this method gives very low values of $N_1$ and $S$ compared to the experiments. The fitting methods give significantly Weissenberg numbers: the first method (targetting S) gives Wi = 7 for Em3 in the experimental conditions investigated, while the second methods leads to Wi = 0.066.

\begin{figure}[t]
\includegraphics[width=0.45\textwidth]{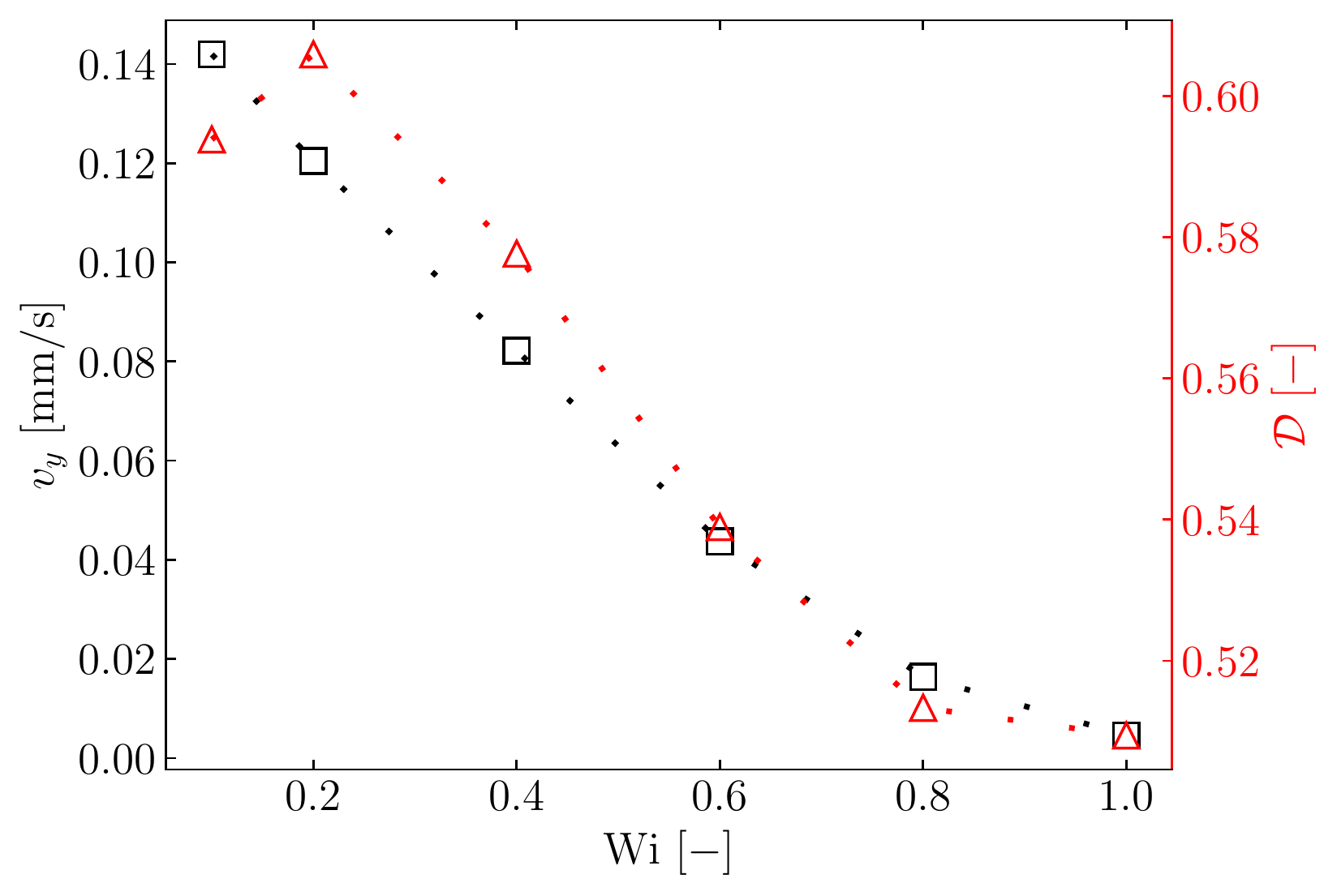}
\centering
\caption{Migration velocity (squares) and Taylor deformation parameter (triangles) of bubble-pair suspended in a VE fluid as a function of $\Wi$. $\Rey = 0.05$, $\Ca = 0.5$. Bubbles are initially aligned in the vorticity direction and $v_y$ is computed at $\mathrm{d}y = 1.25d$.}
\label{fig:migvel_WiSweep}
\end{figure}

To understand further the difference between both fitting methods, we have focused on the effect of elasticity, quantified by the Weissenberg number, on the bubble migration velocity. In Fig. \ref{fig:migvel_WiSweep}, the migration velocity as a function of the Weissenberg number is presented for a repulsive case (the bubbles are aligned exactly in the vorticity direction \ie $\mathrm{d}x = 0$). This shows that the migration velocity $v_y$ decreases with increasing $\Wi$, until not migration occurs at all when Wi = 1. The same trend has been observed for both repulsive and attractive configurations. This means that elasticity does not promote the migration. Thus, it is likely the effect of the Weissenberg number being much higher for the emulsions ($\Wi = 3 \to 9$) which leads to the absence of bubble migration in simulated emulsions with the parameter fitting method based on the elastic number S. This result can be compared with literature studies on droplet migration in viscoelastic fluids. A viscous drop placed near a wall in a Newtonian continuous phase migrates away from the wall \cite{goldsmith1963flow}. Mukherjee \& Sarkar \cite{mukherjee2013effects} numerically studied the effect of continuous phase viscoelasticity on this migration and found that it has a hindering effect. This arises from the difference in the curvature of the streamlines on either side of the drop: they are more curved on the side away from the wall than the side which is confined by the wall, resulting in a force on the drop towards the wall from hoop tensile elastic stresses from the continuous phase, analogous to the Weissenberg effect. A similar effect may play a role here in hindering the lateral bubble pair migration in emulsion simulations using the EVP model at higher $\Wi$ - the interactions between the bubbles distort the streamlines differently in the region of fluid between the bubbles compared to the region of fluid on the "outer" of either bubble, as shown in \figref{fig:oil_rep_intfEvna}. The large dependence of the migration velocity on $\Wi$ may also explain why even in the simulations of the silicone oils, the velocities obtained experimentally (Fig. \ref{Fig_Contour_all}(b)) and numerically (Fig. \ref{fig:vel_transition_numeric}) are the same order of magnitude but not equal.
Additionally, we have quantified the bubble deformation as a function of $\Wi$ in \figref{fig:migvel_WiSweep} with the Taylor deformation parameter, $\mathcal{D} = (L-B)/(L+B)$ where $L$ and $B$ are the major and minor axes of the equivalent ellipsoid in the bubble's middle plane in the $y$-direction, respectively. It can be seen that the relationship is non-monotonic, but at $\Wi\geq 0.4$, $\mathcal{D}$ decreases with increasing $\Wi$, hence the altered bubble deformation with increasing elasticity may have an additional effect on the decreased migration which was observed for the higher $\Wi$ emulsion cases.

Besides, the elongational viscosity is a critical parameter for the characterisation of a viscoelastic material \cite{boger1996viscoelastic}, particularly in the presence of elongational flows \eg as occurs near the droplet tips \cite{aggarwal2008effects}, and thus we may wonder if insufficient characterisation of elongational flow parameters may have led to a discrepancy in the physics modelled. Indeed, the Saramito Herschel-Bulkley shows some distinct properties in extension: for $n\geq 0.5$ where it shows regimes of extension-rate thinning followed by constant extensional viscosity or extensional rate-thickening, which the microstructure of the emulsions cannot support. Simulations with $n=0.4$ were performed in which the Saramito Herschel-Bulkley model shows monotone extension-rate thinning; however, this did not change any of the results qualitatively.

\section{Conclusion}

Bubbles suspended in emulsions tend to align in the direction of the flow under simple shear. The formed patterns look at first sight similar to the particle chains that form in viscoelastic fluids. Two final configurations are obtained: chains of bubbles in contact and alignments of inter-bubble distance larger than the diameter. In the case of chaining, the amount of bubbles in chains depends essentially on the shear strain, while it is also affected by the bubble size and the rheological properties of the emulsion.

Our experimental and numerical investigations of bubble pairs highlight some bubble-specific observations. First, experiments reveal that the interaction between the bubbles is not always attractive: repulsion occurs when the bubbles are aligned in the vorticity direction. In addition, we observe the same interaction pattern in elastoviscoplastic emulsions as in sightly elastic silicone oils. The migration velocities seem to depend on the effective viscosity of the fluid rather than on the elastic stresses.

The simulations successfully reproduce the experimental observations, where bubbles repel when they are aligned in the vorticity direction and attract otherwise. In addition they show that bubble interaction is mainly driven by their deformability, as evidenced by simulations with a Newtonian suspending fluid. In fact, the relative-position-dependent migration occurs only when the capillary number is large enough ($\Ca > 0.2$), so deformation is not negligible. By analysing the pressure fields from the simulations, we note that the capillarity creates a heterogeneous pressure field around each deformed bubble. We show that the cause of bubble migration is the pressure imbalance around each bubble due to the interaction of both heterogeneous pressure fields.
 Moreover, increasing the Weissenberg number with the Oldroyd-B and Saramito-Herschel Bulkley constitutive models hinders bubble migration: the migration velocity decreases when Wi increases from 0 to 1, and larger values of the Weissenberg number ($\Wi = 1 \to 9$) quench the interactions.

Our results show that no alignment criterion can be defined for bubbles based only on the rheological properties of the fluid. Bubble chaining depends mainly on their deformability, and is affected by the fluid viscoelastic properties.

\section*{Acknowledgements}

BF, AJ and AC acknowledge financial support from the Research Council of Norway, within the PIRE project "Multi-scale, Multi-phase Phenomena in Complex Fluids for the Energy Industries", Award Number 1743794. OT and TI received funding from the European Research Council (ERC) under the European Union’s Horizon 2020 research and innovation programme
(Starting Grant StG MUCUS, No. 852529), and from the Swedish Research Council grant VR2017-0489. Computing time was provided by the Swedish National Infrastructure for Computing (SNIC), specifically the PDC Center for High Performance Computing at KTH, the National Supercomputer Centre (NSC) at Link{\"o}ping University, and the High Performance Computing Center North (HPC2N) at Ume{\aa} University.

\bibliographystyle{unsrt}  
\bibliography{Bibliography}

\end{document}